\documentclass[preprints,review,accept,oneauthor,pdftex]{Definitions/mdpi}
\firstpage{1}
\pubvolume{1}
\issuenum{1}
\articlenumber{0}
\pubyear{2021}
\copyrightyear{2020}
\datereceived{}
\dateaccepted{}
\datepublished{}
\hreflink{https://doi.org/} 

\Title{Space-Based Photometry Of Binary Stars: From Voyager To TESS}

\TitleCitation{Space-Based Photometry Of Binary Stars: From Voyager To TESS}

\Author{John Southworth$^{1}$\orcidA{}}

\AuthorNames{John Southworth}

\AuthorCitation{Southworth, J.}

\address{$^{1}$ \quad Astrophysics Group, Keele University, Staffordshire, ST5 5BG, UK; astro.js@keele.ac.uk}

\corres{Correspondence: taylorsouthworth@gmail.com}

\abstract{Binary stars are crucial laboratories for stellar physics, so have been photometric targets for space missions beginning with the very first orbiting telescope (OAO-2) launched in 1968. This review traces the binary stars observed and the scientific results obtained from the early days of ultraviolet missions (OAO-2, Voyager, ANS, IUE), through a period of diversification (\textit{Hipparcos}, WIRE, MOST, BRITE), to the current era of large planetary transit surveys (\textit{CoRoT}, \textit{Kepler}, TESS). In this time observations have been obtained of detached, semi-detached and contact binaries containing dwarfs, sub-giants, giants, supergiants, white dwarfs, planets, neutron stars and accretion discs. Recent missions have found a huge variety of objects such as pulsating stars in eclipsing binaries, multi-eclipsers, heartbeat stars and binaries hosting transiting planets. Particular attention is paid to eclipsing binaries, because they are staggeringly useful, and to the NASA Transiting Exoplanet Survey Satellite (TESS) because its huge sky coverage enables a wide range of scientific investigations with unprecedented ease. These results are placed into context, future missions are discussed, and a list of important science goals is presented.}


\newcommand{\Teff}{\ensuremath{T_{\rm eff}}}                      
\newcommand{\Msun}{\ensuremath{\,{\rm M}_\odot}}                  
\newcommand{\Rsun}{\ensuremath{\,{\rm R}_\odot}}                  

\begin{document} 

\section{Introduction}

In 1923 Hermann Oberth \cite{Oberth23book} pointed out that telescopes placed above Earth's atmosphere would have major advantages versus those that remain on the ground. These include avoiding the blurring effects of the atmosphere and being able to observe at wavelengths at which the atmosphere is opaque. To this we can add the avoidance of scintillation, which causes noise in astronomical photometry \cite{Young67aj,Osborn+15mn}, the ability to observe during daytime, no need to worry about bad weather \cite{Lund20xxx} or adverse ground conditions\footnote{More than one ground-based telescope has been permanently shut down by a lightning strike, and Mt.\ Stromlo Observatory was destroyed by a bushfire.}, and isolation from wildlife\footnote{Ground-based telescopes have been temporarily closed for, among other things, ant infestations, venomous snakes, and mice making bedding out of the cladding of fibre optic cables.}. Siting a telescope in space therefore enables one to obtain data of higher quality, with minimal interruptions, and at wavelengths inaccessible from the ground.

There are obvious disadvantages of space telescopes, including the cost and complexity of constructing them and putting them in orbit, and the inability to perform on-site maintenance and technical interventions. Nevertheless, space-based telescopes can be the best tool with which to tackle a range of scientific goals. This is particularly true for observing programs that require long-term and/or high-precision photometry of a large number of stars. This review is aimed at summarising past work on binary stars using photometry from space telescopes, with a particular emphasis on eclipsing binaries (EBs) and the scientific results coming from the Transiting Exoplanet Survey Satellite (TESS) mission \cite{Ricker+15jatis}.

\subsection{Importance of binary stars}               

The study of EBs is a crucial component in our understanding of stellar physics. It is possible to measure the masses and radii of the components of an EB from directly observed quantities [light curves of eclipses and radial velocity (RV) measurements from spectra] using only algebra \cite{Russell12apj,Kopal59book}, and this has been achieved to high precision (1\% or better) in many cases \cite{Andersen91aarv,Torres++10aarv}. Measurements of the effective temperatures (\Teff s) of the components allow their luminosities to be calculated directly from the well-known equation\footnote{$L$ is luminosity, $R$ is radius and $\sigma$ is the Stefan-Boltzmann constant.} $L = 4 \pi R^2 \sigma {\Teff}^4$, and the inclusion of measured apparent magnitudes of the system enable its distance to be determined in several ways (e.g.\ \cite{Me++05aa}).

Of particular interest are EBs with well-detached (i.e. well-separated) components, as these can be assumed to have evolved as single stars. They can therefore be used to check and calibrate theoretical models of stellar structure and evolution, as the models can be confronted with two stars which have different masses, radii and \Teff s (all measured to 1\% or better) but the same age and chemical composition. Past work has considered phenomena such as the primordial helium abundance \cite{PaczynskiSienkiewicz84apj,Metcalfe+96apj}, the helium-to-metals enrichment ratio \cite{Ribas+00mn}, the strength of convective core overshooting and its dependence on stellar mass \cite{Andersen++90apj,Ribas++00mn,ClaretTorres18apj}, the mixing length parameter $\alpha_{\rm MLT}$ \cite{Graczyk+16aa}, and mass loss by stellar winds \cite{Higl+18aa}. EBs have also been used to establish empirical mass--radius--\Teff--luminosity relations \cite{Torres++10aarv,Moya+18apjs,Eker+18mn} which are useful in particular for measuring the properties of extrasolar planets \cite{Enoch+10aa,Me11mn}. A consistent recent finding is that massive stars have a larger convective core and stronger mixing than predicted by models \cite{Tkachenko+20aa,Martinet+21aa,Johnston21aa}, with rotational mixing an important part of this \cite{Costa+19mn}.

In semi-detached and contact binaries one or both components fills their Roche lobe and mass can be transferred between them or lost from the system. These objects are laboratories for tidal effects, case A mass transfer (during the main sequence), case B mass transfer (during evolution to the giant phase), mass loss, and orbital angular momentum loss (e.g.\ \cite{Dervisoglu+18mn}). When mass is stripped from the originally more massive star it reveals the deep interior and gives a window onto thermonuclear fusion in stars \cite{SarnaDegreve96qjras,Ferraro+06apj,Kolbas+14mn}. Case C mass transfer happens during the evolution of the primary component to the supergiant phase and leads to a wide variety of exotic objects, such as cataclysmic variables (CVs), X-ray binaries, novae, supernovae, millisecond pulsars \cite{LorimerKramer12book}, double-degenerate systems, short gamma-ray bursts \cite{Podsiadlowski+04apj} and gravitational wave events \cite{Belczynski+20aa}.

\subsection{Space photometry and binary stars}        

\begin{figure}[t]
\includegraphics[width=13.5cm]{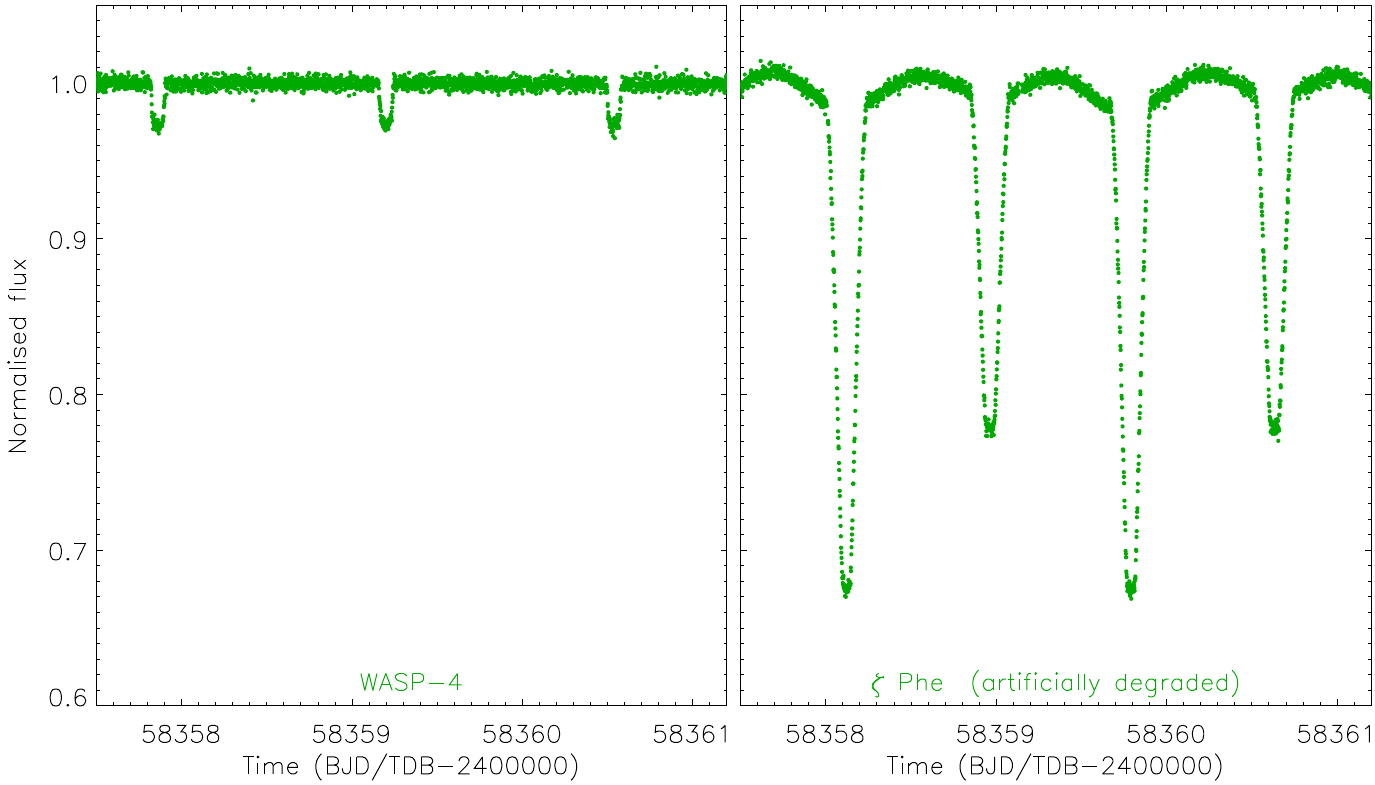}
\caption{Comparison of the brightness variation of a transiting planet (left) and an EB (right), both observed by TESS in sector 2.
WASP-4 was chosen as it shows one of the deepest planetary transits known. $\zeta$\,Phe has a typical brightness variation for an 
EB and was observed by TESS at the same time as WASP-4. The light curve of $\zeta$\,Phe (brightness $V=4.01$) has been artificially
degraded so its scatter matches that of the much fainter WASP-4 system ($V=12.46$). The axes are the same on both plots. Only a
small fraction of the light curves are shown. \label{fig:compare}}
\end{figure}

In 1946 Lyman Spitzer \cite{Spitzer46report} discussed in detail the establishment of optical and ultraviolet (UV) telescopes in space. One of the science cases for this was the measurement of the bulk and atmospheric properties of EBs. This vision was realised in the 1970s and 1980s with many small UV telescopes placed in orbit (see reviews by Wilson \& Boksenberg \cite{WilsonBoksenberg69araa} and Savage \cite{Savage99aspc}), several of which were used to study bright detached, semi-detached, and contact binaries as well as evolved objects such as CVs (e.g.\ OAO-2, ANS, IUE). The available instrumentation broadened hugely in the 1990s and 2000s with the establishment of larger general-purpose observatories (e.g.\ HST), single-project telescopes (e.g.\ \textit{Hipparcos}) and a varied range of smaller missions (e.g.\ WIRE and MOST). More recently, the era of space-based planetary transit searches has brought many observations of a huge number of point sources (\textit{CoRoT}, \textit{Kepler}, TESS) and -- \emph{as a byproduct} -- has revolutionised studies of binary stars.

A particular feature of transit search telescopes is the long duration and high precision of the photometry they obtain, mandated by the goal to find the smallest possible planets\footnote{Planetary transit depths scale as the square of the ratio of the radius of the planet to the star, so small planets produce very small transits.} with the largest possible orbital periods $P$. Any mission capable of finding planet transits is also able to obtain high-quality photometry for EBs due to their (usually) much larger eclipse depths. To illustrate this point, Fig.\,\ref{fig:compare} compares the light curves of WASP-4 (a planetary system with one of the deepest transits known) and $\zeta$\,Phe (a typical bright EB) obtained simultaneously by TESS.

EBs can, in turn, be used as an independent check on the timestamps from space satellites (Von Essen et al.\ \cite{Vonessen+20aj}).

TESS is the most important of these missions because its huge sky coverage means it is obtaining light curves of the vast majority of the bright EBs known, making it easy to pick particular targets of interest with little advance planning and no need to visit a telescope. One downside is that the coarse angular resolution (a TESS pixel subtends 21$^{\prime\prime}$$\times$21$^{\prime\prime}$ and its angular resolution is approximately 1$^{\prime}$) means the light curves of many point sources are contaminated with light from nearby objects. This is particularly important for short-period binaries as almost all have a fainter companion (Tokovinin et al.\ \cite{Tokovinin+06aa}).

\subsection{Contents of this review}                  

This review presents a detailed history of the use of space-based light curves in binary star science. It begins with a brief look at each of the early missions, then summarises the results from the \textit{Kepler} and TESS satellites in detail. It concludes with a recap, a look at future space missions, a consideration of the current status of space observations for binary stars, and a list of future work to be tackled.

The review contains some discussion of all important space missions and picks out the best scientific highlights. No attempt is made to summarise every mission ever flown, or every scientific result ever published, as it would have become a book. Details of those missions that are covered are shown in Table\,\ref{tab:missions}.

\begin{specialtable}[H] 
\caption{Basic information on the space telescopes discussed in this work. \label{tab:missions}}
\tablesize{\footnotesize} 
\setlength{\tabcolsep}{5pt}
\begin{tabular}{lllccl}
\toprule
\textbf{Mission / }  & \textbf{Launch date} & \textbf{End date} & \textbf{Wavelength} & \textbf{Aperture} & \textbf{Sect.} \\
\textbf{instrument}  &                      &                   & \textbf{range (nm)} & \textbf{(cm)}     &                \\
\midrule                                                                                                                   
OAO                  & 1968 Dec 7           & 1973 Jan          & 105--420            & 20--40            & 2.1            \\
ANS                  & 1974 Aug 30          & 1976 Jun          & 155--329            & 22                & 2.2            \\
Voyager 1 / UVS      & 1977 Sep 5           & ongoing           & 50--170             & 4                 & 2.3            \\
Voyager 2 / UVS      & 1977 Aug 20          & ongoing           & 50--170             & 4                 & 2.3            \\
IUE                  & 1978 Jan 26          & 1996 Sep 30       & 115--330            & 45                & 2.4            \\
\textit{Hipparcos}   & 1989 Aug 8           & 1993 Aug 15       & ~~~380--630 $^*$    & 29                & 2.5            \\
WIRE                 & 1999 Mar 5           & 2006 Oct 23       & $\sim$$V$+$R$       & 5.2               & 2.6            \\
MOST                 & 2003 Jue 30          & 2019 Mar          & ~~~400--700 $^*$    & 15                & 2.7            \\
\textit{CoRoT}       & 2006 Dec 27          & 2012 Oct          & ~~~450--850 $^*$    & 27                & 2.8            \\
BRITE                & 2013 to 2014         & ongoing           & 400--700            & 3                 & 2.9            \\
HST                  & 1990 Apr 24          & ongoing           & UV--opt--IR         & 240               & 2.10           \\
\textit{Spitzer}     & 2003 Aug 25          & 2020 Jan 30       & IR                  & 85                & 2.10           \\
INTEGRAL / OMC       & 2002 Oct 17          & ongoing           & $V$-band            & 5                 & 2.10           \\
Coriolis / SMEI      & 2003 Jan 6           & ongoing           & 450--950            & $\approx$1        & 2.10           \\
STEREO A / HI-1      & 2006 Oct 26          & ongoing           & 630--730            & 1.6               & 2.10           \\
STEREO B / HI-1      & 2006 Oct 26          & 2014 Oct 1        & 630--730            & 1.6               & 2.10           \\
LUT                  & 2013 Dec 1           & ongoing           & 245--340            & 15                & 2.10           \\
\textit{Kepler} + K2 & 2009 Mar 7           & 2018 Oct 30       & ~~~440--840 $^*$    & 95                & 3              \\
TESS                 & 2018 Apr 18          & ongoing           & ~~~590--990 $^*$    & 10.5              & 4              \\
\textit{Gaia}        & 2013 Dec 19          & ongoing           & 320--1000           & 145 $\times$ 50   & 5.2            \\
CHEOPS               & 2019 Dec 18          & ongoing           & ~~~400--800 $^*$    & 32                & 5.2            \\
PLATO                & 2026 planned         &                   & $\sim$ 500--900     & $\sim$12          & 5.2            \\
\bottomrule
\end{tabular}
\newline $^*$ Approximate wavelengths of half the peak response, as estimated from passband response functions.
\end{specialtable}


\section{Early missions}

\subsection{Orbiting Astronomical Observatory}        

\begin{figure}[t]
\includegraphics[width=13.5cm]{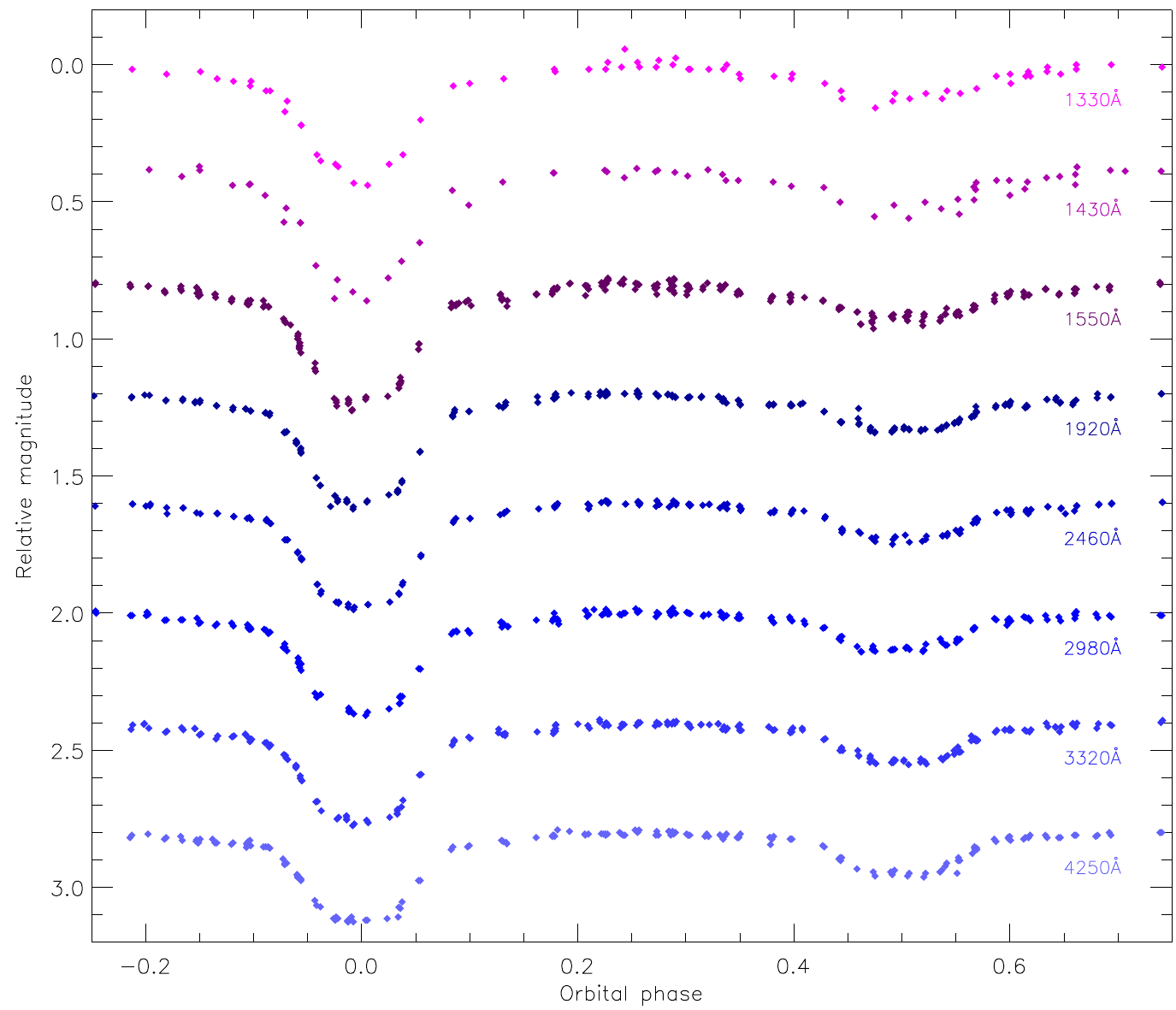}
\caption{The OAO-2 satellite light curve of the EB VV\,Ori ($P = 1.485$\,d) in seven UV passbands
(labelled), published by Eaton \cite{Eaton75apj}. \label{fig:vvori}}
\end{figure}

The era of space photometry started as long ago as the 1960s, with early missions concentrating on the UV wavelengths that find Earth's atmosphere impassable. This naturally led to observations of high-energy phenomena such as hot stars and mass transfer within binary systems.

The Orbiting Astronomical Observatory 2 (OAO-2) was the first operational space-based telescope. It was launched by NASA in 1968, after the failure of OAO-1 in 1966, and was active until 1973. It carried several UV telescopes in the Wisconsin Experiment Package (Code et al.\ \cite{Code+70apj}) to obtain UV photometry and scanning spectroscopy. An extensive program of observations included the early-type EBs U\,Oph \cite{EatonWard73apj}, LY\,Aur \cite{Heap73apj}, and VV\,Ori (see Fig.\,\ref{fig:vvori}) and MR\,Cyg \cite{Eaton75apj}, with findings generally in agreement with those from ground-based data. The data on U\,Oph confirmed that its gravity darkening follows theoretical expectations \cite{Vonzeipel24mn} and allowed its limb darkening to be measured in the UV. Eaton \cite{Eaton75pasp} obtained a light curve of the prototypical semi-detached eclipsing binary $\beta$\,Per.

Also observed were $\beta$\,Lyr \cite{Kondo++76apss}, the Wolf-Rayet + O\,star system V444\,Cyg \cite{Cherepashchuk++84apj} and the B-type + supergiant systems 31\,Cyg and 32\,Cyg \cite{Doherty++74apj,DohertyJung75apj}. These observations allowed the density profiles in the atmospheres of the Wolf-Rayet and supergiant stars to be measured. Other targets included the classical nova FH\,Ser \cite{GallagherCode74apj}, enabling the tracing of its luminosity after eruption, and the symbiotic system AG\,Peg \cite{Gallagher+79apj} which contains a Wolf-Rayet star and an M3 giant filling its Roche lobe.

There was also an OAO-3 satellite, called \textit{Copernicus} (Rogerson et al.\ \cite{Rogerson+73apj}) which was very similar to OAO-2. A small amount of spectroscopic work on binary systems was done with this satellite, e.g. for $\beta$\,Lyr (Hack et al.\ \cite{Hack+74nat}) and $\gamma$\,Vel (St.-Louis et al.\ \cite{Stlouis++93apj}).

\subsection{Astronomical Netherlands Satellite}       

The Astronomical Netherlands Satellite (Astronomische Nederlandse Satelliet, ANS) was launched by NASA in 1974 and observed until 1976. The Ultraviolet Experiment Package on the ANS was a 22\,cm telescope with a spectrophotometer observing simultaneously in five passbands from 155\,nm to 329\,nm (Van Duinen et al.\ \cite{Vanduinen+75aa}). Notable was the concentration on CVs, including the dwarf novae U\,Gem and VW\,Hyi in quiescence \cite{WuPanek82apj} and outburst \cite{WuPanek83apj}, V1500\,Cyg 100\,d after a nova event \cite{WuKester77aa}, and the low-mass X-ray binaries HZ\,Her (Her X-1) and V1357\,Cyg (Cyg X-1) \cite{Wu+82pasp}.

Less extreme binaries were also observed during the short operational duration of the ANS. The high-mass binary $\delta$\,Pic was studied by Eaton \& Wu \cite{EatonWu83pasp}, who found that one or both components might be intrinsically variable and that there was no evidence for circumstellar matter, but were not able to conclusively decide whether it was a detached or semi-detached binary. Kondo et al.\ \cite{Kondo++81apj} observed 14 close binaries and found a strong UV excess in four of them. The RS\,CVn system U\,Cep was observed by Kondo et al.\ \cite{Kondo++78apj}, who inferred the presence of a hot spot on each star due to gas streams in the system. The contact binaries W\,UMa and 44\,Boo (i\,Boo) were observed by Eaton \& Wu \cite{EatonWu81aj}, who found that the former has UV properties consistent with its optical ones but that the latter is better explained by the presence of a dark spot on its surface. Eaton et al.\ \cite{Eaton++80apj} observed W\,UMa and concluded that at UV wavelengths it shows limb darkening but very little gravity darkening.

\subsection{Voyager}                                  

The two NASA \textit{Voyager} missions were launched in 1977 to take advantage of a rare celestial opportunity that would enable them to visit all four of the gas-giant planets in the Solar system. Both host a UV spectrometer (UVS) that used an objective grating to obtain spectra covering 50--170\,nm (Broadfoot et al.\ \cite{Broadfoot+77ssrv}). These were intended for studying the atmospheric structures and compositions of the atmospheres of the giant planets in our Solar system, but were also useful for observations of UV-bright stars. Although both \textit{Voyager} spacecraft remain operable, the UVSs are no longer in use.

Spectroscopic light curves were obtained by one or both \textit{Voyagers} for EBs such as $\beta$\,Lyr in 1983-5 (Kondo et al.\ \cite{Kondo+94apj}) and V356\,Sgr in 1986-7 (Polidan \cite{Polidan89ssrv}), allowing the circumstellar material in these systems to be probed. \textit{Voyager} UVS was also used to study some contact binaries but the complexity of the data and their interpretation meant few results were published.

\subsection{International Ultraviolet Explorer}       

The International Ultraviolet Explorer (IUE; Boggess et al. \cite{Boggess+78nat}) was the first satellite which made a wide-ranging impact on astrophysics. Launched in 1978 by NASA, it was operational for 18 years until being shut down in 1996 for financial reasons. During this time it obtained over 100\,000 UV spectra of objects ranging from solar system bodies to active galaxies. It was equipped with two instruments, the Short Wavelength Spectrograph (SWS) and the Long Wavelength Spectrograph (LWS) that could each be operated in two resolution modes. Spectrophotometry could be obtained using either spectrograph with the use of the larger aperture, which could then be divided up into light curves with arbitrary passbands. The high-resolution spectra were also useful for radial velocity studies (e.g.\ Stickland \& Lloyd \cite{SticklandLloyd01obs} and references therein).

Po{\l}ubek \cite{Polubek03adspr} has catalogued the light curves of EBs obtained from IUE, comprising 9889 observations of 328 targets. Whilst most have relatively few data, 30 objects were observed at least 100 times. Within this category are $\beta$\,Per \cite{Brandi+89aa}, the prototype of the EA class of eclipsing binary, and $\beta$\,Lyr \cite{Kondo+94apj}, the prototypical EB-type eclipsing binary. Observations of these and other interacting binaries have helped us to understand the mass loss and its kinematics in these systems plus the prevalence and effects of colliding stellar winds \cite{Sahade94apss}.

\subsection{Hipparcos}                                

\begin{figure}[t]
\includegraphics[width=13.5cm]{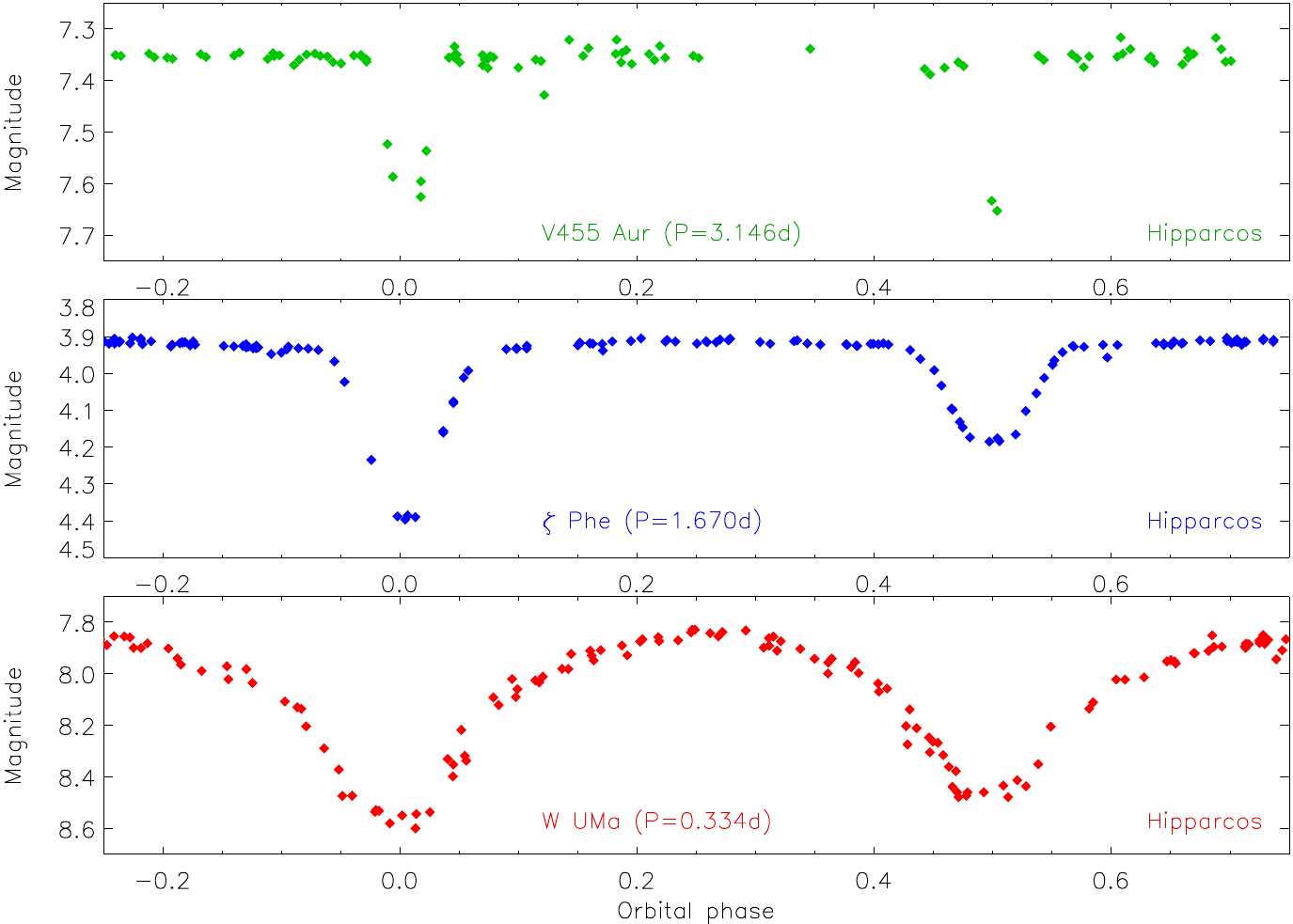}
\caption{The \textit{Hipparcos} satellite light curves of three example EBs. V455\,Aur (top) was chosen
as it was these data that enabled its discovery. $\zeta$\,Phe (middle) was selected for comparison with
the TESS data in Fig.\,\ref{fig:compare}. W\,UMa (bottom) was picked as it is the prototype eclipsing contact
binary; the TESS data of this object are shown in Fig.\,\ref{fig:sd+c} for comparison. \label{fig:hip}}
\end{figure}

The ESA \textit{Hipparcos} mission measured the trigonometric parallaxes of over 100\,000 nearby stars \cite{Perryman+97aa} over the time period 1989--1993. As part of this process it collected light curves with an average of 110 epochs per star \cite{Vanleeuwen+97aa}. As well as providing new data for known systems, this dataset led to the discovery of 343 new EBs. These were given designations in the General Catalogue of Variable Stars by Kazarovets et al.\ \cite{Kazarovets+99ibvs}.

The existence of direct distance measurements for EBs containing stars of known radius means they could be used to calibrate the radiative flux scale. Popper \cite{Popper98pasp} did this for spectral types B6--F8 using 14 EBs with precise \textit{Hipparcos} parallaxes. Ribas et al.\ \cite{Ribas+98aa} used \textit{Hipparcos} parallaxes to determine \Teff s of 20 EBs and compare them to \Teff s from photometric calibrations, finding a good agreement albeit with the parallax-derived \Teff s being slightly smaller. Kruszewski \& Semeniuk \cite{KruszewskiSemeniuk99aca} identified 198 EBs with good parallaxes that could be used to calibrate the relation between surface brightness and colour. Semeniuk \cite{Semeniuk00aca} compared \textit{Hipparcos} and photometric distances for 19 EBs and found a good agreement.

An example of the usefulness of the \textit{Hipparcos} light curves is that they led to the discovery of eclipses in HD\,45191, subsequently given the designation V455\,Aur (see Fig.\,\ref{fig:hip}). This prompted Griffin \cite{Griffin01obs,Griffin13obs} to obtain a high-quality spectroscopic orbit, and then Southworth \cite{Me21obs4} to determine the physical properties of the system to high precision using the light curve of this object from the TESS satellite. \textit{Hipparcos} light curves of several other EBs have been used to mimic the expected properties of the photometry from the \textit{Gaia} satellite to explore the impact of \textit{Gaia} on the study of EBs \cite{Munari+01aa,Zwitter+03aa}.

\subsection{Wide Field Infrared Explorer}             
\label{sec:wire}

\begin{figure}[t]
\includegraphics[width=13.5cm]{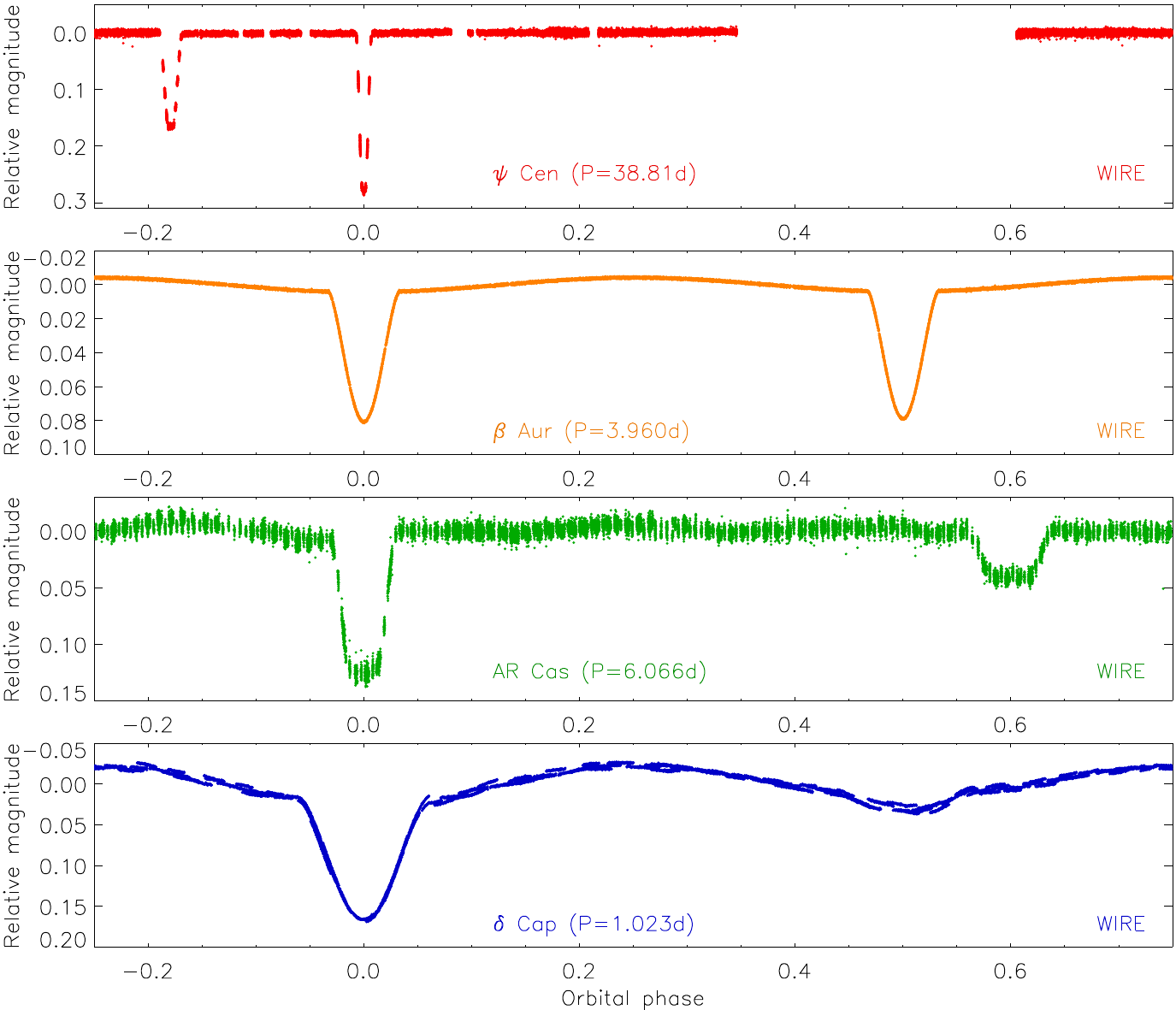}
\caption{The WIRE satellite light curves of four EBs (labelled)
plotted according to orbital phase. \label{fig:wire}}
\end{figure}

The Wide Field Infrared Explorer (WIRE) satellite was launched by NASA in 1999 to obtain infrared (IR) images of galaxies showing active star formation. A flaw in the electronics caused the loss of hydrogen coolant for the IR camera, which outgassed and spun up the satellite. By the time it was brought back under control, there was no coolant left and thus the main mission could not be achieved. The star tracker was subsequently used from 2003-6 as a high-precision space-based photometer. WIRE had been used to observe about 300 stars \cite{Bruntt07coast} before contact was lost in 2008 \cite{BrunttSouthworth08conf}.

Although the main focus of WIRE was on asteroseismology, it was the first instrument to obtain light curves of such high precision from space and represents the start of the era of space-based high-precision photometry of binary stars. The first result was the discovery that the bright star ($V=4.03$) $\psi$\,Cen is an EB (Bruntt et al.\ \cite{Bruntt+06aa}). When combined with data from the Solar Mass Ejection Imager (SMEI) it was possible to measure the orbital period of the system (38.81\,d) plus the photometric parameters of the eclipses. The second result was the first good light curve of $\beta$\,Aur (Southworth et al.\ \cite{Me++07aa}), one of the brightest known EBs ($V=1.90$), the first binary with a double-lined spectroscopic orbit (Baker \cite{Baker10}) and the first detached EB for which the radii of the stars were measured directly (Stebbins \cite{Stebbins11apj}).

A total of seven EBs were observed using WIRE. Two more for which the light curves are valuable (but remain under analysis) are AR\,Cas, which shows total eclipses and slowly-pulsating B-star (SPB) pulsations, and $\delta$\,Cap, a short-period system with $\gamma$\,Dor pulsations. The WIRE light curves for all four EBs are shown in Fig.\,\ref{fig:wire}. Much of the scatter for AR\,Cas and $\delta$\,Cap are due to the pulsations, which are not commensurate with the orbital period.

\subsection{MOST}                                     

The Microvariability and Oscillations of Stars (MOST) satellite was launched by the Canadian Space Agency in 2003 and functioned as a small space-based photometer until completion of the mission in 2019 \cite{Walker+03pasp}. It was primarily operated as a single-object photometer for asteroseismic studies, but observed a small number of EBs for various science goals.

HD\,313926 was discovered to be a short-period EB (2.27\,d) containing two B stars in an eccentric orbit when it was used as a guide star for MOST (Rucinski et al.\ \cite{Rucinski+07mn}). The light curve is of high quality but spectroscopic observations have not yet been published; follow-up of this object will allow the physical properties of two young B stars to be determined.

Pribulla et al.\ \cite{Pribulla+08mn} observed a set of blue stragglers in the open cluster M67 using MOST, with the aim of testing the hypothesis that blue stragglers primarily form through binary interactions \cite{Bailyn95araa}. Excellent light curves were obtained for the contact binaries AH\,Cnc, ES\,Cnc and EV\,Cnc, and two more EBs were discovered among the non-members of the cluster. Pribulla et al.\ \cite{Pribulla+10an} presented a catalogue of 16 new and four known EBs observed using MOST, many of which were observed as guide stars.

The EB system $\mu$\,Eri, containing a pulsating star, was observed using MOST (Jerzy- kiewicz et al.\ \cite{Jerzykiewicz+13mn}). A total of 31 frequencies were measured, two eclipses were analysed, and the physical properties of the primary star determined. The frequencies were inferred to be of the SPB type. Another pulsating system observed using MOST is $\delta$\,Ori\,A, which was found to show tidally perturbed g-mode oscillations at 12 different frequencies in the primary star (Pablo et al.\ \cite{Pablo+15apj}).

Windemuth et al.\ \cite{Windemuth+13apj} used MOST to observe the perplexing EB $\theta^1$\,Ori\,B$_1$, also called BM\,Ori. This consists of a B3 star in orbit with a larger but less massive component which is surmised to be a self-luminous disc-shaped object. The MOST data showed that the eclipses had changed shape compared to previous observations, with a much shorter duration of totality. BM\,Ori may be an extremely young system with one component still in the protostar phase.

Nemravov\'a et al.\ \cite{Nemravova+16aa} obtained a MOST light curve of the quadruple system $\xi$\,Tau, which contains a 7.14\,d-period EB with third and fourth components on 145\,d and 51\,yr orbits. Augmented with ground-based photometry, spectroscopy and interferometry, the physical properties of the component stars were established. Non-Keplerian motion was detected in the form of apsidal motion and eclipse timing variations. A quasi-periodic brightness modulation was found and attributed to surface spots or pulsations on the third component in the system.

\subsection{CoRoT}                                    

\begin{figure}[t]
\includegraphics[width=13.5cm]{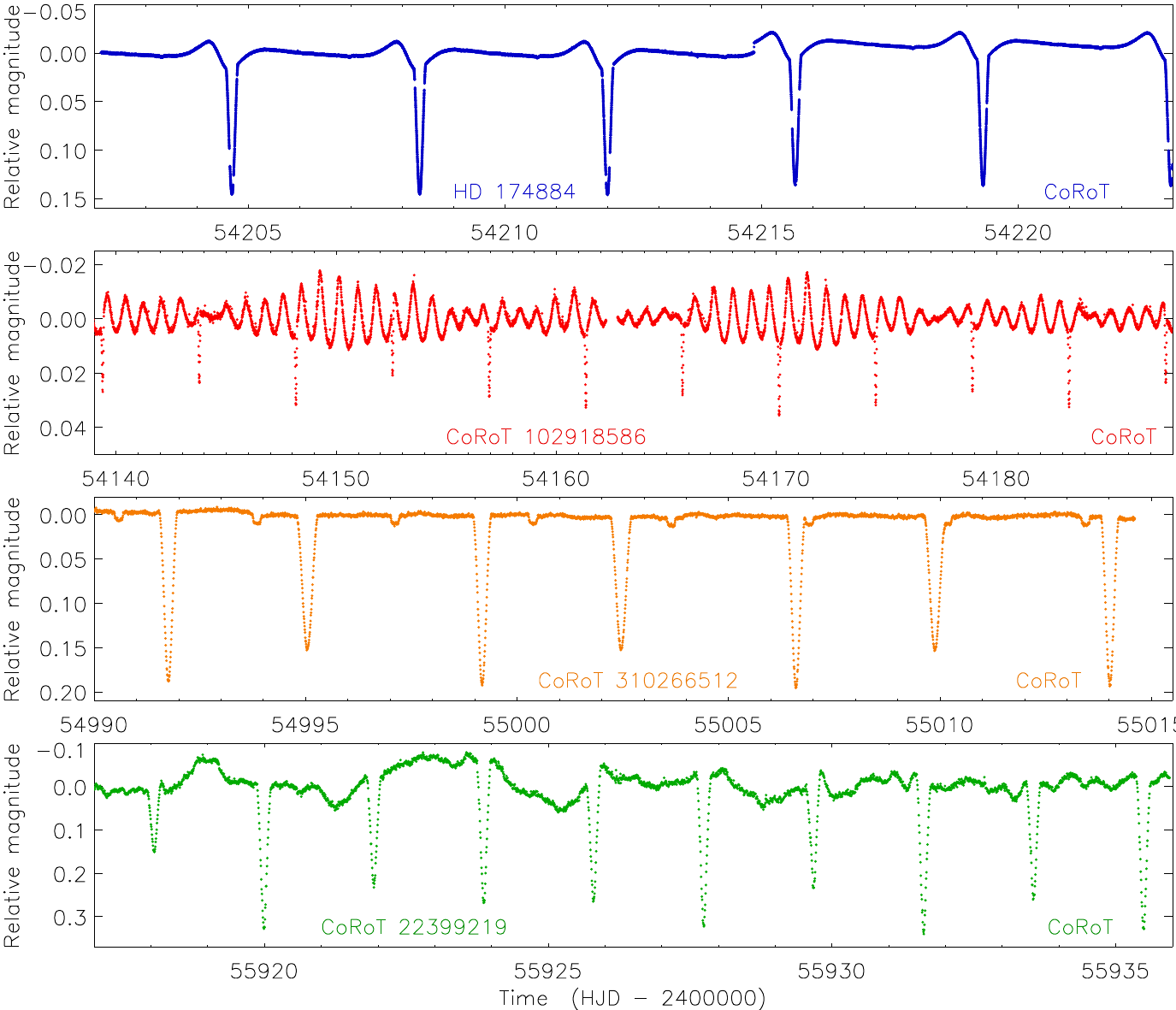}
\caption{Parts of the \textit{CoRoT} satellite light curves of four EBs (labelled). \label{fig:corot}}
\end{figure}

The \textit{Convection, Rotation et Transits plan\'etaires (CoRoT)} satellite was the first space platform to obtain well-sampled light curves of a large number of stars \cite{Auvergne+09aa,Baglin+16book}. The telescope on \textit{CoRoT} consisted of a 27\,cm diameter off-axis mirror feeding two channels, each with two CCDs. The asteroseismology channel was slightly defocussed in order to provide high-precision photometry for bright stars (see also \cite{Me+09mn}). The exoplanet channel was designed to observe a larger number of fainter stars; a prism was stationed before the CCD to allow three-colour light curves to be obtained for brighter objects in the field. It was led by the French National Centre for Space Studies (CNES), launched into a low-Earth polar orbit in December 2006, and continued until its failure in October 2012.

\textit{CoRoT} performed observations of 25 fields, with durations from 21 to 152\,d, towards the Galactic centre and anti-centre. It observed a total of 163\,665 stars primarily for performing asteroseismology and searching for transiting planets. A significant number of these were inevitably known or unknown EBs. Deleuil et al.\ \cite{Deleuil+18aa} catalogued 2269 EBs detected in the exoplanet channel plus 584 candidate or confirmed transiting planets.

An early EB result from \textit{CoRoT} was the discovery of eclipses in HD\,174884 (Macer- oni et al.\ \cite{Maceroni+09aa}), a short-period (3.657\,d) eccentric ($e=0.29$) double-lined binary (Fig.\,\ref{fig:corot}). The primary and secondary eclipses are of very different depth (140\,mmag and 1.5\,mmag) due partly to the orientation of the eccentric orbit. The properties of the component stars are consistent with it being a young system ($\sim$125\,Myr), in agreement with its eccentric orbit. More intriguingly, seven short-period oscillations were detected, of which six are at multiples of the orbital frequency $f_{\rm orb}$, the strongest being at $8f_{\rm orb}$ and $13f_{\rm orb}$. These are consistent with being tidally excited pulsations.

A range of other EBs containing pulsating stars was also detected, including \textit{CoRoT} 102931335 and \textit{CoRoT} 102918586 which show $\gamma$\,Dor pulsations \cite{Damiani+10apss,Maceroni+13aa} (Fig.\,\ref{fig:corot}), the oscillating Algol system (oEA; \cite{Mkrtichian+02aspc,Mkrtichian+03aspc}) \textit{CoRoT} 102980178 \cite{Sokolovsky+10coast}, \textit{CoRoT} 105906206 which shows total eclipses and 50 $\delta$\,Scuti oscillation frequencies \cite{Dasilva+14aa}, and \textit{CoRoT} 100866999 for which 89 $\gamma$\,Dor and 35 $\delta$\,Sct oscillation frequencies were measured \cite{ChapellierMathias13aa}. HSS\,348 is a spectroscopic binary of spectral type B8\,+\,B8.5 that was found to show shallow eclipses in its \textit{CoRoT} light curve \cite{Strassmeier+17aa}. It also exhibits rotational modulation due to dark spots on one or both components, which may be related to the fact that at least one is a HgMn (chemically-peculiar) star.

A set of EBs was found which masqueraded as transiting planets and therefore were studied in detail. The detection and rejection of false positives is an inescapable part of wide-field surveys for transiting planets, and EBs blended with a brighter object are capable of mimicking a transiting planet signal extremely well (e.g.\ \cite{Mandushev+05apj,Odonovan+06apj2}). The planet candidate \textit{CoRoT} 101186644 was established to be an M-dwarf transiting a late-F star (Tal-Or et al.\ \cite{Talor+13aa}) because the velocity amplitude of the primary star was too high for its companion to be planetary; \textit{CoRoT} 110680825 was rejected as a planetary candidate when a faint second set of lines were found in its spectrum (Tal-Or et al.\ \cite{Talor+11aa}) -- it turned out to be a binary of an A-type dwarf and an F-type dwarf showing grazing eclipses that were diluted by the light from a late-G giant. On this theme, Fernandez \& Chou \cite{FernandezChou15pasp} found and studied \textit{CoRoT} 310266512, an EB that shows primary and secondary eclipses on a period of 7.421\,d and tertiary eclipses on a 3.266\,d period (Fig.\,\ref{fig:corot}).

Circumbinary material has been seen in the \textit{CoRoT} light curves of at least two binary systems. AU\,Mon is a 11.11-d period semi-detached binary composed of a Be star primary and a G-type giant filling its Roche lobe. Desmet et al.\ \cite{Desmet+10mn} determined the physical properties of the system and attributed the 417\,d cyclic variation to brightness variations in the circumbinary disc. Gillen et al.\ \cite{Gillen+14aa} found that \textit{CoRoT} 223992193 is a pre-main-sequence star embedded in a circumbinary disc (Fig.\,\ref{fig:corot}). Their precise measurements of the masses and radii of the stars are consistent with an age of only 5\,Myr for this object, which is a member of the young open cluster NGC\,2264.


\subsection{BRITE}                                    

BRITE-Constellation consists of five\footnote{Six BRITE satellites were launched but one never functioned.} nanosatellites in low-Earth orbit which obtain wide-field photometry of very bright stars \cite{Weiss+14pasp,Pablo+16pasp}. Each has a 3\,cm diameter telescope and observes through either a blue (400--450\,nm) filter (BRITE-Austria and BRITE-Lem) or a red (550--700\,nm) filter (UniBRITE, BRITE-Heweliusz and BRITE-Toronto). Multiple satellites observe the same field simultaneously or in series to obtain multi-colour photometry over long time intervals, and 25 fields have been observed so far (Weiss et al.\ \cite{Weiss+21uni}). The restriction to bright apparent magnitudes (mostly $V < 6$) means the mission concentrates on the brightest stars such as giants and hot dwarfs.

A highlight of BRITE observations so far is an explanation of the photometric variations of the Wolf-Rayet binary system EZ\,CMa (Schmutz \& Koenigsberger \cite{SchmutzKoenigsberger19aa}), which required continuous observations over a period of a few weeks. This system shows variability on a 3.7\,d period due to the increase in brightness of the shock from the wind of the Wolf-Rayet star around periastron passage, overlaid on eclipses of the shocked region. The crucial finding is that the system shows a small eccentricity ($e=0.10 \pm 0.01$) and an extraordinarily fast apsidal period of $U = 100 \pm 5$\,d. The latter causes changes in the effects of binarity over timescales of months, confounding studies based on data covering a longer timespan. BRITE has also been used to observe other binary systems such as the eclipsers $\beta$\,Lyr, $\beta$\,Aur, $\zeta$\,Aur and $\epsilon$\,Aur and the non-eclipsing RS\,CVn-type system V711\,Tau, $\nu$\,Cen and $\gamma$\,Lup \cite{Rucinski+18aj,Rucinski+19aj,Strassmeier+20aa,Jerzykiewicz+21mn}.

Another bright binary observed using BRITE is the Wolf-Rayet system $\gamma^2$\,Vel, for which Richardson \cite{Richardson+17mn} found the photometric variations to be caused by collision of the winds of the component stars. The prototypical mysterious binary $\eta$\,Car was also observed, with the BRITE data suggesting the presence of tidally-induced pulsations (Richardson et al.\ \cite{Richardson+18mn}). Pablo et al.\ \cite{Pablo+17mn,Pablo+19mn} have found the massive binaries $\iota$\,Ori and $\epsilon$\,Lup to show weak ``heartbeat'' signatures in their BRITE light curves due to the strengthening of the ellipsoidal and reflection effects near periastron. Jerzykiewicz et al.\ \cite{Jerzykiewicz+20mn} observed $\pi^5$\,Ori and found it to be an ellipsoidal variable with SPB pulsations.


\subsection{Other space missions}                     

For reasons of ``space'' a few other missions are discussed only briefly. The Hubble Space Telescope (HST) is the best-known, but is a general-purpose observatory whose time is extremely valuable. It has thus not been used much for the time-intensive work of obtaining light curves of EBs. Jeffers et al.\ \cite{Jeffers+06mn,Jeffers+06mn2} used the Space Telescope Imaging Spectrograph (STIS) and G430L grism to obtain spectrophotometry of the active binary system SV\,Cam -- they used eclipse mapping to determine the starspot distribution on the surface of the primary component and studied the wavelength dependence of its limb darkening. The Fine Guidance Sensors (FGS) have been used to obtain precise parallaxes and astrometry of binary stars, leading to many mass measurements (see e.g.\ Benedict et al.\ \cite{Benedict+17pasp}). Edmonds et al.\ \cite{Edmonds+96apj} used WFPC to monitor the central region of the globular cluster 47\,Tuc, finding eight binary stars of a range of types. Massey et al.\ \cite{Massey++02apj} obtained photometry and spectroscopy with STIS of four high-mass spectroscopic binaries, of which three are eclipsing, in the R136 cluster in the LMC; full light curves of these objects have yet to be obtained due to the difficulty of obtaining light curves in such a crowded field. Bonanos et al.\ \cite{Bonanos+19aa} have made a comprehensive study of variability in HST observations and presented the Hubble Catalog of Variables (HCV).

The \textit{Spitzer} Space Telescope is another of NASA's Great Observatories that was able to perform long-term photometric monitoring but was not widely used for binary star work. Gillen et al.\ \cite{Gillen+20mn} used \textit{CoRoT} to find and \textit{Spitzer} to characterise MON-735 a pre-main-sequence EB containing two 0.3\Msun\ stars, situated in the young open cluster NGC\,2264.

The ESA INTErnational Gamma-Ray Astrophysics Laboratory (INTEGRAL) satellite was launched in October 2002 and remains operational at the time of writing. It uses three instruments to observe photons of energies from 3\,keV to 10\,MeV, augmented by an Optical Monitoring Camera (OMC) to provide simultaneous optical observations of the targets \cite{Winkler+03aa,Kuulkers+21newar}. OMC is a 50\,mm diameter refractor with a $V$ filter and a 5$^\circ$$\times$5$^\circ$ field of view. This large field means it has obtained light curves of many EBs. Zasche \cite{Zasche08newar,Zasche09newar,Zasche10newar,Zasche11newar} has presented results on 77 EBs observed using INTEGRAL/OMC.

The Coriolis satellite was launched by the US Naval and Air Force Research Laboratories (NRL and AFRL) in January 2003 to obtain imaging of Earth and of the Sun. The latter is performed by the Solar Mass Ejection Imager (SMEI), which consists of three CCDs observing a 3$^\circ$$\times$160$^\circ$ strip of sky starting near the Sun and ending at the anti-Sun direction \cite{Jackson+04soph}. Hounsell et al.\ \cite{Hounsell+1apj} presented a catalogue of nova eruption light curves from SMEI observations. The SMEI data of $\psi$\,Cen were crucial for determining the orbital period of this EB, as the WIRE light curve (see Section\,\ref{sec:wire}) covers only one primary and one secondary eclipse (Bruntt et al.\ \cite{Bruntt+06aa}).

Two Solar Terrestrial Relations Observatory (STEREO) satellites were launched by NASA in October 2006 to obtain stereoscopic imaging of the Sun. Both were equipped with, among other things, heliospheric imagers that observed along the ecliptic plane \cite{Eyles+09soph}. Wraight et al.\ \cite{Wraight+11mn} presented observations of 263 EBs brighter than magnitude 10.5, and Wraight et al.\ \cite{Wraight+12mn} found nine systems containing at least one low-mass star.

The following mission \emph{is} a telescope above Earth's atmosphere, but has a very different platform to the others discussed in this review. The Lunar-based Ultraviolet Telescope (LUT) is a 15\,cm reflecting telescope mounted on the Chang'e-3 Moon lander \cite{Cao+11scpma}. It was launched by the China National Space Administration in December 2013 and remains operational on the Moon's surface. The LUT uses a CCD to observe a 1.36$^\circ$$\times$1.36$^\circ$ field of view with a passband covering 245--345\,nm \cite{Wang+15raa,Meng+17aspc}. It has been used to obtain light curves of several binary systems including the semi-detached binaries AI\,Dra, TW\,Dra and V548\,Cyg \cite{Liao+15apss,Liao+16raa,Zhu+16aj}, and the contact binary V921\,Her \cite{Zhou+16adast}.


\section{Kepler and K2}

The {\it Kepler} mission proved to be the first of a new generation of space telescopes, as its observations greatly exceeded previous missions in both quality and duration. {\it Kepler} was a 0.95\,m diameter Schmidt telescope launched in March 2009 into an orbit trailing the Earth around the Sun. Science observations began in May 2009 and concentrated on a single field of size 105\,deg$^2$ in Lyra and Cygnus. {\it Kepler} simultaneously observed up to 512 targets at short cadence (58.8\,s) and up to 170\,000 targets at long cadence (1765.5\,s) (Borucki \cite{Borucki16rpph}). Observations continued until May 2013, when the failure of a second of its four reaction wheels meant the telescope could no longer point accurately enough for its science goals.

After extensive experimentation the spacecraft was brought back into service for a revised mission dubbed K2 (Howell et al.\ \cite{Howell+14pasp}), beginning in March 2014 and ending in September 2018 when its fuel supply ran out. K2 was restricted to observing near the ecliptic plane in order to use solar radiation pressure to compensate for the loss of fine pointing, and observed 19 fields (``campaigns'') before its decommissioning.

{\it Kepler} was conceived to find transiting extrasolar planets (Borucki et al.\ \cite{Borucki+11apj}) and was extremely successful \cite{Rowe+14apj,Marcy+14apjs,Morton+16apj}. Observations could be requested for other science cases through several mechanisms. A large number of EBs were observed by {\it Kepler}, either on request or discovered from the {\it Kepler} light curves. The Kepler Eclipsing Binary Working Group catalogued a total of 2878 EBs observed by {\it Kepler} \cite{Prsa+11aj,Kirk+16aj}, as well as classifying their ``detachedness'' \cite{Matijevic+12aj} and weeding out false positives \cite{Abdulmasih+16aj}.

\begin{figure}[t]
\includegraphics[width=11cm]{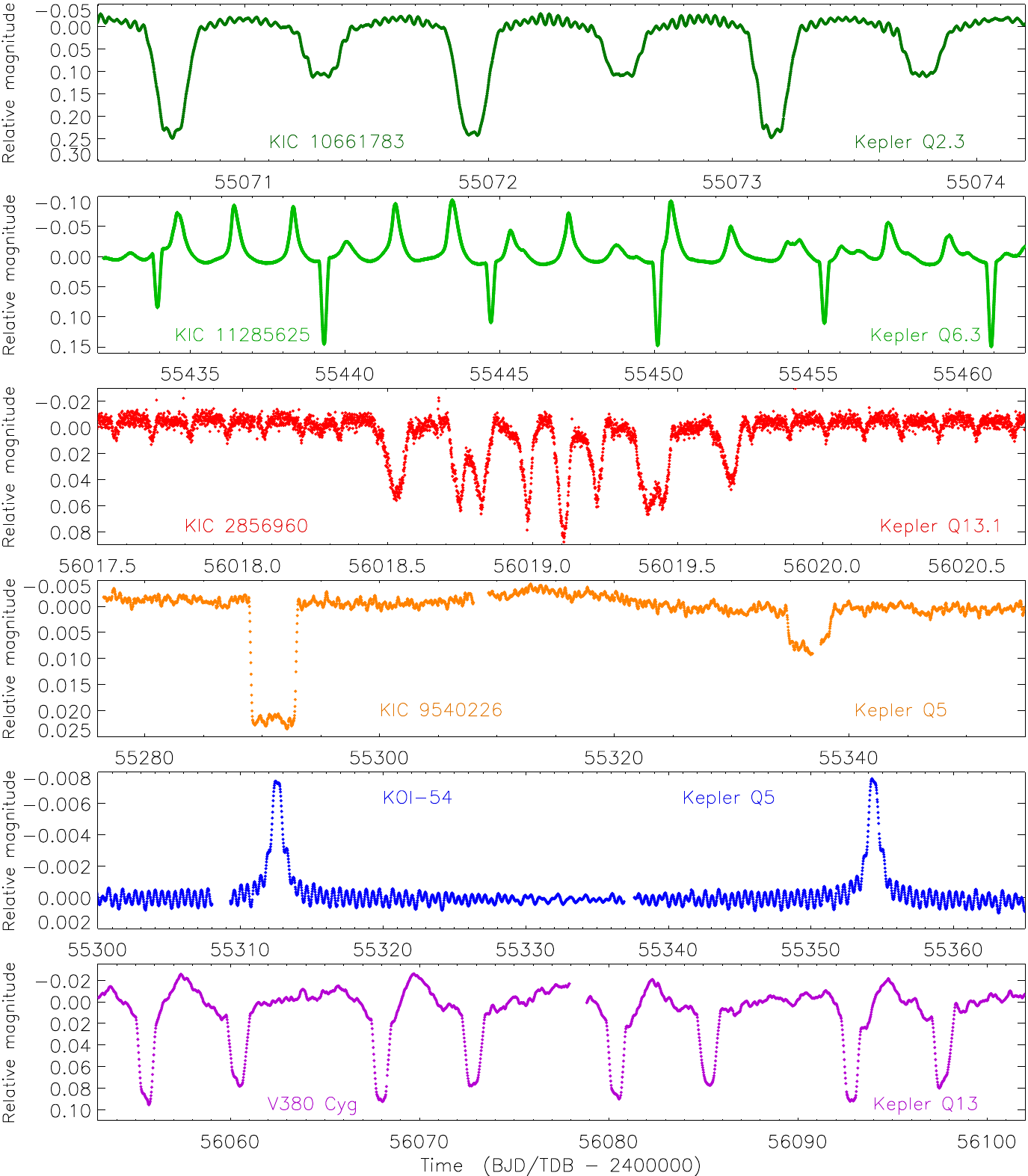}
\caption{Example \textit{Kepler} satellite light curves of a set of binary systems. The names and the \textit{Kepler} quarters are labelled.
\label{fig:kepler}}
\end{figure}

\subsection{``Normal'' EBs}                           

The {\it Kepler} database has been valuable for studying exotic objects, but equally important for providing observations of previously unobtainable quality for more straightforward EBs in pursuit of the science goals outlined in Section 1. Several groups have embarked on spectroscopic follow-up of a sample of (typically bright) {\it Kepler} EBs. Examples include 20 objects from a set of 75 defined by He{\l}miniak et al.\ \cite{Helminiak+16mn,Helminiak+17mn,Helminiak+19mn}, 43 long-period EBs from K2 campaigns 1--3 for which tidal effects are negligible (Maxted \& Hutcheon \cite{MaxtedHutcheon18aa}), and the SDSS-HET survey (Mahadevan et al.\ \cite{Mahadevan+19aj}).

One area where significant effort has been made is EBs containing low-mass stars. There remains a ``radius inflation'' problem for stars below 1\Msun\ in EBs, where theoretical models underpredict the measured radii by up to 15\% and overpredict their \Teff s. It has been suggested that this is due to tidal effects causing a faster rotation and thus a stronger magnetic field \cite{Hoxie73aa,Ribas06apss,Lopez07apj}, but this conflicts with the fact that it has also been seen in single stars \cite{Berger+06apj,Spada+13apj,MorrellNaylor19mn}, so there is no current consensus on the cause \cite{Torres13an}. One way to investigate this is to study EBs containing low-mass stars, preferably with a range of mass, metallicity and rotational velocity. {\it Kepler} has contributed to this work with light curves of low-mass EBs such as KIC\,6131659 (Bass et al.\ \cite{Bass+12apj}), KIC\,1571511 (Ofir et al.\ \cite{Ofir+12mn}), T-Cyg1-12664 (Han et al.\ \cite{Han+17aj}) and 231 objects identified by Coughlin et al.\ \cite{Coughlin+11aj}.

Another area where EBs can be used to test stellar theory in detail is to study those that are members of stellar clusters, because theoretical models must then match the properties of the EB as well as the other cluster members for a single age and chemical composition \cite{Me++04mn,Me++05aa,Brogaard+11aa,Brogaard+12aa}. {\it Kepler} did not make a major contribution here because only four stellar clusters were observed within the {\it Kepler} field, although the data did allow detailed investigations of the open clusters NGC\,6819 \cite{Brewer+16aj}, NGC\,6811 \cite{Sandquist+16apj} and NGC\,6791 \cite{Yakut+15mn}. The K2 mission has observed many more clusters and further results are expected. Torres et al.\ \cite{Torres+18apj,Torres+19apj,Torres+20apj} studied three EBs in Ruprecht 147, determining the properties of the component stars plus the age, metallicity and distance of the cluster.

\subsection{Giant systems}                            

One type of EB which particularly benefited from {\it Kepler} observations is systems containing giant stars. The large sizes of giants means that detached systems must have orbital periods of tens of days or more and thus long eclipses that are very difficult to properly observe from the ground. Many giants exhibit solar-like oscillations with periods long enough to be studied with long-cadence data, although these are suppressed by tidal effects for shorter-period systems. Gaulme et al.\ \cite{Gaulme+16apj} found that the oscillations were undetectable in {\it Kepler} data for systems with $\frac{R_{\rm giant}+R_{\rm dwarf}}{a} \geqslant 0.16$, where $R_{\rm giant}$ and $R_{\rm dwarf}$ are the radii of the giant and its companion and $a$ is the semimajor axis of the relative orbit.

The first giant EB to be found with {\it Kepler} was KIC\,8410637 (Hekker et al.\ \cite{Hekker+10apj}), which showed one eclipse in the {\it Kepler} Q0 (quarter zero) light curve. Frandsen et al.\ \cite{Frandsen+13aa} used additional {\it Kepler} data and ground-based spectroscopy to determine its orbital period (408\,d) and the properties of the component stars. Revised properties of this and three others were presented by Theme{\ss}l et al.\ \cite{Themessl+18mn} and used to check whether the asteroseismic scaling relations \cite{KjeldsenBedding95aa} predict masses and radii in agreement with those directly measured for the eclipsing giants. The \textit{Kepler} light curve is shown in Fig.\,\ref{fig:kepler}, in which the primary eclipse (an occultation), secondary eclipse (a transit), solar-like oscillations from the giant, and an increased brightness at periastron (around time JD 2455313) can be seen.

Other giants in EBs have been studied by Gaulme et al.\ \cite{Gaulme+13apj,Gaulme+14apj,Gaulme+16apj}. He{\l}miniak et al.\ \cite{Helminiak+15apj} and Rawls et al.\ \cite{Rawls+16apj} measured the properties of the double-red-giant EB KIC\,9246715 and found that asteroseismic scaling relations correctly predicted the measured properties of the oscillating secondary component. Li et al.\ \cite{Li+18mn} used six EBs with giant components to calibrate the convective mixing-length parameter ($\alpha_{\rm MLT}$), finding it to be larger for giant stars than the Sun. Benbakoura et al.\ \cite{Benbakoura+21aa} used a larger sample of giants in EBs to find that the asteroseismic scaling relations overpredict the giant masses by 15\% and radii by 5\%, but that this could be reduced by recalibrating the relations.

\subsection{Multiple eclipsers}                       
\label{sec:kepler:multi}

Before {\it Kepler} only a few  objects were known which contained more than two types of eclipses. This was primarily due to the lack of long-term high-precision light curves of many stars, as they are intrinsically rare objects often showing long and/or shallow eclipses.

KOI-126 was an early discovery from {\it Kepler}. It consists of an EB of two M dwarfs (masses 0.24\Msun\ and 0.21\Msun) in a 1.77\,d orbit, itself eclipsing and eclipsed by a slightly evolved G star (mass 1.3\Msun\ and radius 2.0\Rsun) on an orbit of period 33.9\,d (Carter et al.\ \cite{Carter+11sci}). The durations and times of the eclipses of the M-dwarfs depend on their instantaneous orbital separation, and on their orbital velocities relative to the wider orbit, so the masses and radii of all three stars could be determined directly from the {\it Kepler} light curve plus RVs of the brighter star.

A second early result from {\it Kepler} was HD\,181068 (``Trinity''), a similar system but scaled up. The third star is a giant (3.0\Msun\ and 12.5\Rsun) in a 45.5\,d orbit with an EB (0.92\Msun\ and 0.87\Msun\ dwarfs) with a period of 0.906\,d. All three stars have similar \Teff s (5100\,K, 5100\,K, 4700\,K) so the rather odd situation appears where lots of short and shallow eclipses occur, much longer eclipses happen on an apparent 25\,d orbit, and the shallow eclipses are visible only during alternate long eclipses. The discovery of the system was announced by Derekas et al.\ \cite{Derekas+11sci}) and a detailed analysis was performed by Borkovits et al.\ \cite{Borkovits+13mn}.

An even more intriguing system is KIC 4150611 (He{\l}miniak et al.\ \cite{Helminiak+17aa}) which shows eclipses on four different periods (1.43, 1.52, 8.65, 94.2 d) plus hybrid $\delta$\,Sct\,/\,$\gamma$\,Dor pulsations. Three objects are seen in high-resolution imaging. The brightest star is an A-type pulsator and shows a small forest of eclipses every 94.2\,d due to being orbited by a 1.52\,d EB containing two K or M stars. The 8.65\,d period is an EB of two G stars and the 1.43\,d period is likely from the third point source which is a background star.

Other multiply-eclipsing systems include KIC 4247791 (Lehmann et al.\ \cite{Lehmann+12aa}) which shows two sets of eclipses plus spectral lines from four stars; KIC\,6543674, which shows eclipses on a 2.39\,d period with extra eclipses from a third body on a 1090-d orbit (Masuda et al.\ \cite{Masuda+15apj}); KIC 7177553 which shows four sets of spectral lines and one set of eclipses (Lehmann et al.\ \cite{Lehmann+16apj}); and the doubly-eclipsing systems EPIC 219217635 and EPIC 249432662 (Borkovits et al.\ \cite{Borkovits+18mn,Borkovits+19mn}) observed during the K2 mission and showing eclipse timing variations due to being hierarchical multiple systems. KIC 2856960 shows eclipses on a 0.258\,d period and forests of eclipses every 204\,d (Fig.\,\ref{fig:kepler}; Armstrong et al.\ \cite{Armstrong+12aa}), similar to KIC 4150611, but its configuration remains unclear as no model has been found that satisfies the observations (Marsh et al.\ \cite{Marsh++14mn}).

\subsection{Planets transiting binary stars}          
\label{sec:kepler:planet}

A special case of hierarchical multiple system is an EB orbited by one or more planets on a P-type orbit (planet orbits both stars in a binary). A total of 14 circumbinary planets are known, in 12 systems, and all were found due to their transits. The unique and unmistakable signature of these objects is eclipses in an EB with additional transits by a smaller object, where the additional transits are much less frequent and show large changes in their timings and durations. The additional transits are shorter when the planet and transited star are orbiting in different directions, and longer when the orbital motion of both objects is in the same direction. The additional transits can occur early or late depending on where the stars are in the EB orbit at the time of transit, and normally occur in pairs as both stars are transited at similar points in each planet orbit.

Kepler-16 was the first known circumbinary planet (Doyle et al.\ \cite{Doyle+11sci}), followed by Kepler-34 and Kepler-35 (Welsh et al.\ \cite{Welsh+12nat}). Kepler-47 is the only system known to host more than one planet, having three with orbital periods of 49.5, 303.2 and 187.4\,d (Orosz et al.\ \cite{Orosz+12sci,Orosz+19aj}). The orbital periods of the known circumbinary planets are all unusually long, ranging from 49.5\,d to 1107\,d, due to the requirement of orbital stability. The large transit timing and duration variations hold information on the positions and velocities of the components, meaning circumbinary planetary systems can be characterised in detail from their light curve and RVs of just one of the component stars.

\subsection{Pulsators}                                

Like \textit{CoRoT}, {\it Kepler} provided previously unobtainable data on a range of binary systems containing pulsating stars. An early result was KIC 10661783 (Fig.\,\ref{fig:kepler}) which was found to be a semi-detached EB with total eclipses and a large number of $\delta$\,Scuti oscillation frequencies (Southworth et al.\ \cite{Me+11mn}). Lehmann et al.\ \cite{Lehmann+13aa} obtained spectroscopy and measured the physical properties of the component stars, finding that the system must be detached and therefore in a post-mass-transfer phase. A comprehensive reanalysis of this system (Miszuda et al.\ \cite{Miszuda+21mn}) found 590 oscillation frequencies including a set at lower frequency that are likely g-mode pulsations and some that may be tidally excited. The properties of the system could only be matched by theoretical models accounting for binarity and with an enhanced opacity at several regions in the primary star. $\delta$\,Scuti oscillations have been found in the {\it Kepler} light curve of other binary systems, e.g.\ KIC 3858884 (Maceroni et al.\ \cite{Maceroni+14aa}), KIC 4544587 (Hambleton et al.\ \cite{Hambleton+13mn}), KIC 10736223 (Chen et al.\ \cite{Chen+20apj}), KIC 4851217 and KIC 10686876 (Liakos et al.\ \cite{Liakos20aa}).

Another class of pulsator for which major results have been obtained from {\it Kepler} is the $\gamma$\,Dor stars. Debosscher et al.\ \cite{Debosscher+13aa} discovered glorious $\gamma$\,Dor pulsations in KIC 11285625 (Fig.\,\ref{fig:kepler}), determined the physical properties from the {\it Kepler} data and ground-based spectroscopy, and found evidence for rotational splitting of the pulsation modes. Guo et al.\ \cite{Guo+16apj} found $\gamma$\,Dor and $\delta$\,Sct pulsations in KIC 9851944 but were not able to place detailed constraints on stellar theory because the masses were measured to only 4\% precision. Lee \cite{Lee16apj} found KIC 6048106 to be a semi-detached system showing $\gamma$\,Dor oscillations. Guo et al.\ \cite{Guo++17apj2} studied KIC 9592855, a post-mass-transfer system with $\gamma$\,Dor and $\delta$\,Sct pulsations in the secondary component that require both its core and envelope to rotate synchronously with the orbital motion. Sekaran et al.\ \cite{Sekaran+20aa,Sekaran+21aa} determined the masses and radii of KIC 9850387 to high precision and measured a large number of p-mode and g-mode frequencies. They found that the asteroseismic results constrained the interior properties of the stars better than their physical properties, that the envelope of the primary star was rotating three times faster than its core, and that interior mixing processes were stronger in its interior than for single stars of its mass and age.

A common, almost ubiquitous, feature of pulsations in close binary stars is tidal excitation and/or perturbation of oscillation modes. A vivid example of this is KOI-54 (Welsh et al.\ \cite{Welsh+11apjs}), a non-eclipsing binary with a 41.8-d eccentric orbit which shows brightenings near periastron and strong oscillations at exactly 90 and 91 times $f_{\rm orb}$, (Fig.\,\ref{fig:kepler}). The brightenings arise from the reflection and ellipsoidal effects in this highly-eccentric ($e=0.83$) and low-inclination ($i=5.5^\circ$) binary. On closer inspection there are oscillations at many integer multiples of $f_{\rm orb}$, from 29 to 91, and these are caused by resonance between the pulsation modes and the dynamical tides at periastron. The striking resemblance of the {\it Kepler} light curve to an echocardiogram trace prompted this type of object to be called a ``heartbeat star'' and over 100 are now known \cite{Thompson+12apj,Beck+14aa,Kirk+16aj,Cheng+20apj}. A theoretical treatment for tidally-excited oscillations has been developed by Fuller \cite{Fuller17mn} and applied to several objects (e.g.\ KIC 3230227 \cite{Guo++17apj}).

The strict periodicity of pulsations means that observations of them are affected by the light-time effect arising from binary orbits. This leads to a new class of binary star system that can be studied using the change in pulsation properties with orbital phase. Such objects can be labelled ``PB1'' if one star pulsates and ``PB2'' if both stars pulsate, where PB stands for pulsating binary, by analogy to the SB1 and SB2 spectroscopic designations. The FM (frequency modulation) method was developed by Shibahashi \& Kurtz \cite{ShibahashiKurtz12mn} and yields information on the RVs of the stars. The PM (phase modulation) method (Murphy et al.\ \cite{Murphy+14mn}) is analogous and gives the change in orbital \emph{position} of the stars. Murphy et al.\ \cite{Murphy+18mn} have used this method to determine orbital parameters for 317 PB1 and 24 PB2 systems. Murphy et al.\ \cite{Murphy+16apj} used the PM method to discover a planet on an 840-d period orbit around a pulsating A star.

Another highlight from {\it Kepler} is the discovery and characterisation of stochastic g-mode oscillations in the massive EB V380\,Cyg (Fig.\,\ref{fig:kepler}; Tkachenko et al.\ \cite{Tkachenko+12mn,Tkachenko+14mn}). Theoretical models were unable to match the observed properties, a discrepancy that can be improved by adding extra mixing to the models. A subsequent determination of the atmospheric parameters of this system \cite{Tkachenko+20aa} removed this discrepancy.

\subsection{Evolved binaries}                         

Aside from the semi-detached binaries discussed in the previous section, and a smattering of papers on contact binaries, {\it Kepler} data have contributed to the understanding of more evolved binaries. Bloemen et al.\ \cite{Bloemen+11mn} analysed the {\it Kepler} light curve of KPD 1946+4340, a binary with a subdwarf B star (0.47\Msun, 0.21\Rsun, $\Teff = 34500$\,K) and a white dwarf (0.59\Msun, 0.014\Rsun, 15900\,K). The light curve was found to show eclipses, reflection and ellipsoidal effects, gravitational lensing and Doppler beaming \cite{ShakuraPostnov87aa,Zucker++07apj}. The measured amplitude of the Doppler beaming was in perfect agreement with the expected value.

Gravitational lensing \cite{Marsh01mn} is expected to be negligible in most binaries but can reach an observable amplitude in systems with a relatively large orbital separation and containing at least one degenerate object. The effect was discovered in KOI-3278, a system with a 0.6\Msun\ white dwarf and a 1.0\Msun\ G dwarf in an 88.1\,d orbit (Kruse \& Agol \cite{KruseAgol14sci}). The result in this case is a primary eclipse (where the G-dwarf occults the white dwarf) of depth 0.1\% and a ``secondary eclipse'' (white dwarf in front of G dwarf) of an \emph{increase} in flux of 0.1\% due to gravitational lensing of the G-dwarf's light by the white dwarf. Three more examples have been presented by Kawahara et al.\ \cite{Kawahara+18aj} and a fifth object by Masuda et al.\ \cite{Masuda+19apj}.

A total of nine known CVs were observed by {\it Kepler} \cite{Howell+13aj}, and a few have been newly discovered or are present as background objects around other {\it Kepler} targets \cite{Brown+15aj}. V447\,Lyr was seen to show eclipses, short outburst and long outbursts \cite{Ramsay+12mn}. V344\,Lyr and V1504\,Cyg have been the subjects of multiple studies \cite{OsakiKato13pasj,Dobrotka+16mn}. A new CV was found during the K2 mission, as a variable background star to the F-dwarf EPIC 203830112 \cite{Riddenharper+19mn}.



\section{TESS}

The Transiting Exoplanet Survey Satellite (TESS) was launched by NASA in April 2018 on a two-year mission to find planets transiting bright stars (Ricker et al.\ \cite{Ricker+15jatis}), and is currently funded until at least September 2022. It consists of four cameras, each with a field of view of $24^\circ\times24^\circ$ and an aperture of diameter 10.5\,cm. The fields of view are arranged into a strip covering $24^\circ\times96^\circ$ at a pixel scale of 21$^{\prime\prime}$\,px$^{-1}$. TESS has a highly elliptical 13.7\,d orbit around Earth and observes a single field for two consecutive orbits; each is called a \emph{sector} and thus lasts 27.4\,d minus time lost for transmitting the data to Earth. The data from TESS are literally falling out of the sky at us, although in practice they are picked up from the ground by NASA's Deep Space Network.

In many ways TESS is a successor to the {\it Kepler} satellite but with a significantly different science goal. {\it Kepler} was designed to find Earth-like planets and thus measure their occurrence rate in the Galactic neighbourhood. The small size of Earth relative to the Sun, and its one-year orbital period, means its transits are extremely shallow (0.008\%) and infrequent. Detecting them thus requires very high-quality observations of many stars for more than three years (so three transits can be seen). {\it Kepler} therefore consisted of a large telescope (which in turn meant a small field of view and thus the need to observe relatively faint stars) pointed at a single field for its entire mission lifetime (which dictated a heliocentric orbit). This caused three issues: only a small fraction of the known variable stars were observed by {\it Kepler}; these tended to be relatively faint and thus time-consuming to follow up with spectroscopy; and telemetry limitations meant that the data for only a small fraction of the stars observed were actually sent to Earth.

The main science goal of TESS is to find a large number of planets orbiting bright stars. It therefore consists of multiple cameras observing much brighter stars for a shorter period of time in order to cover almost the entire sky. A wonderful byproduct of this is that it has observed many binary systems that are well-known and have long observational histories. An important advantage of its orbit is that much higher data rates are possible, so observations of every star seen by the TESS CCDs are sent to Earth with at least one sample every 1800\,s \footnote{The TESS full-frame images originally had a 1800\,s cadence but this was subsequently decreased to 600\,s.}. A typical workflow for {\it Kepler} data was to find an EB of interest and obtain spectroscopic observations to study it in detail. With TESS it is now possible to select EBs for which good spectroscopy is already available, download the light curve from the MAST archive\footnote{\texttt{https://mast.stsci.edu/portal/Mashup/Clients/Mast/Portal.html}}, and go straight into the analysis without expending any time or effort in obtaining follow-up observations. There are, of course, exceptions in both cases but it is clear that TESS makes researchers' lives much easier.

Whilst the TESS mission is ongoing, several surveys for EBs have already been published in pursuit of different science goals. Justesen \& Albrecht \cite{JustesenAlbrecht21apj} compiled a catalogue of EBs observed by TESS in sectors 1--13 and used their measured orbital periods and eccentricities to study tidal effects in binary stars. Whilst cooler stars were found to behave as predicted by dynamical-tide theories, binaries containing hotter stars (6250 to 10000 K) have lower eccentricities than expected. This indicates that either tidal circularisation is more efficient for these stars than expected, or that their formation and pre-main-sequence evolution have an important effect on their eccentricities. IJspeert et al.\ \cite{IJspeert+21aa} compiled an homogeneous sample of OBA-type EBs with a particular aim of facilitating asteroseismology of high-mass stars in EBs. Comprehensive catalogues of EBs observed by TESS are currently in preparation by at least two groups.

\subsection{Standard EBs}                             

\begin{figure}[t]
\includegraphics[width=13.5cm]{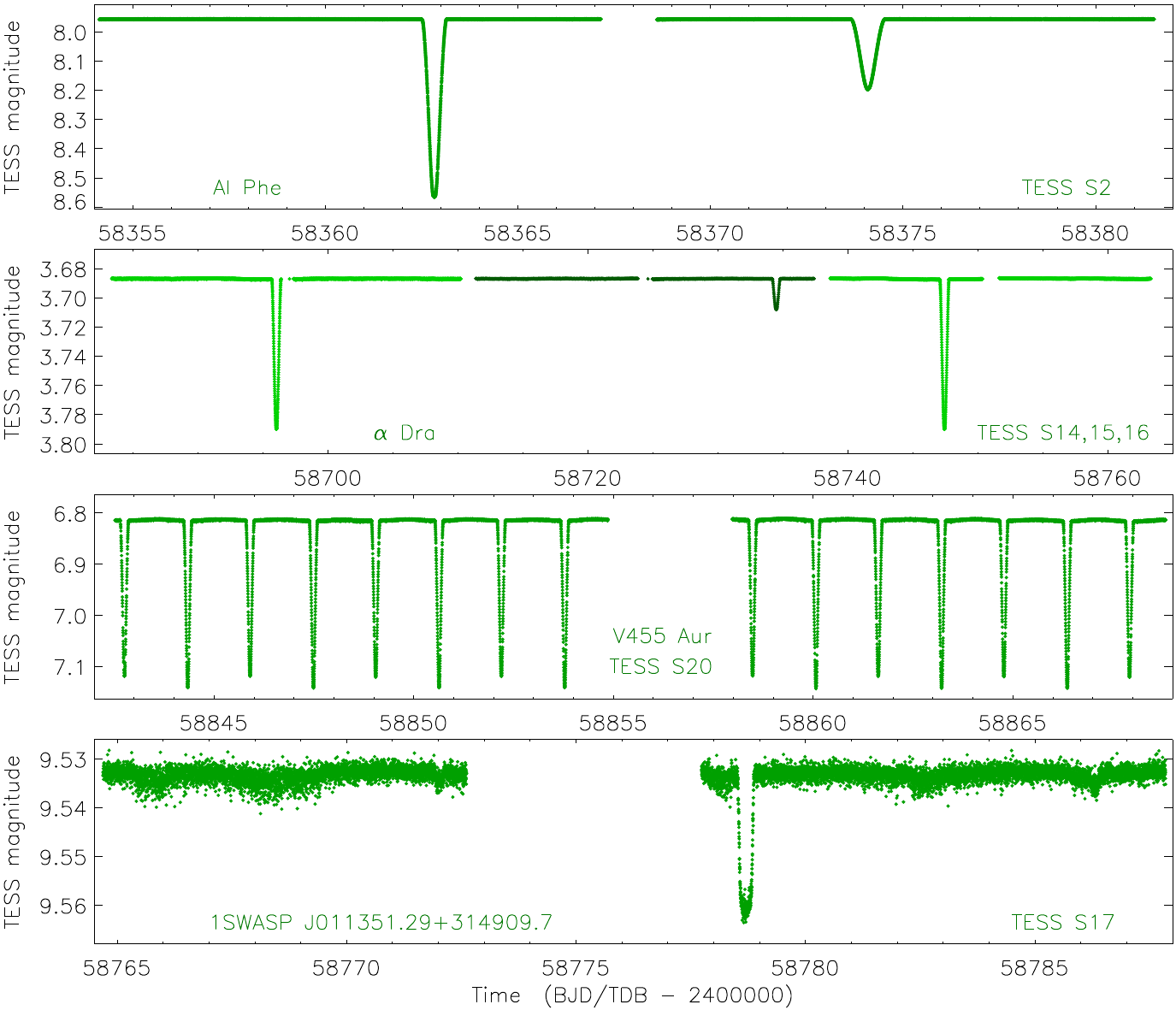}
\caption{Example TESS light curves of binary systems. The names and sectors are labelled. The data
for $\alpha$\,Dra have been binned to decrease the size of the image file. \label{fig:normal}}
\end{figure}

TESS observations are suitable for finding eclipses in some of the brightest stars in the sky. This is particularly useful as they are usually too bright to observe easily from the ground due to CCD saturation and the lack of nearby comparison stars. Bedding et al.\ \cite{Bedding++19rnaas} announced the discovery of eclipses in $\alpha$\,Dra ($V = 3.60$) and Hey et al.\ (submitted) have presented a detailed analysis of the system. It shows shallow eclipses (9\% and 2\% depths) and has an eccentric orbit ($e=0.42$) with a period of 51.4\,d (Fig.\,\ref{fig:normal}). The TESS light curves of $\beta$\,Aur ($V=1.90$) and $\beta$\,Per ($V=2.12$) are also useful, although the eclipse depths are unreliable for stars as bright as this.

Another well-known EB is AI\,Phe, which has total eclipses, a long observational history and an exceptionally good TESS light curve (Fig.\,\ref{fig:normal}). Maxted et al.\ \cite{Maxted+20mn} organised the independent analysis of these data by multiple researchers in order to see how well they agreed and thus to what level of precision it is possible to determine the radii of stars in EBs. From 13 different solutions of two types of TESS data by nine researchers using five different light curve synthesis codes, it was found that the measured fractional radii (the radii of the stars expressed as a fraction of the semimajor axis of the relative orbit) were consistent at the 0.1\% level. Thus it is currently possible to measure the radii of well-behaved EBs to 0.1\% given good data. Maxted et al.\ included high-quality published RVs of the stars to determine the masses and radii to the highest precision ever achieved for an EB: $1.1938 \pm 0.0008$\Msun\ and $1.8050 \pm 0.0022$\Rsun\ for the primary star and $1.2438 \pm 0.0008$\Msun\ and $2.9332 \pm 0.0023$\Rsun\ for the more evolved secondary star.

There are many other well-known EBs for which precise results have been obtained from TESS data. AL\,Dor and BN\,Scl were studied by Graczyk et al.\ \cite{Graczyk+21aa} as part of a systematic recalibration of the surface brightness relations for dwarf stars. HS\,Hya used to show eclipses that were deeper than 0.5\,mag in the 1970s (Gyldenkerne et al.\ \cite{Gyldenkerne++75aa}), around 0.05\,mag deep in the 2000s (Zasche \& Paschke \cite{ZaschePaschke12aa}), and are now on the verge of disappearing entirely (Davenport et al.\ \cite{Davenport+21aj}). A systematic reanalysis by the current author has begun in order to curate the DEBCat catalogue of EBs with masses and radii measured to 2\% or better \cite{Me15debcat}. This has so far led to revised and improved properties for $\psi$\,Cen \cite{Me20obs}, KX\,Cnc \cite{Me21obs1}, V1022\,Cas \cite{Me21obs2}, AN Cam \cite{Me21obs3}, V455\,Aur \cite{Me21obs4} (Fig.\,\ref{fig:normal}) and V505\,Per \cite{Me21obs5} with more results in the pipeline. Zasche et al.\ \cite{Zasche+20aa} have used TESS data, among other sources, to detect the light-time effect in 14 EBs.

TESS can also contribute to period analysis for EBs with apsidal motion. Baroch et al.\ \cite{Baroch+21aa} used TESS data to study apsidal motion in 15 EBs and determine the apsidal periods of nine of these. A comparison of these with theoretical models by Claret et al.\ \cite{Claret+21aa} found good agreement.

TESS preferentially observes bright stars, so low-mass stars are under-represented in this sample. However, a significant number of low-mass EBs have been observed. Miller et al.\ \cite{Miller+21pasp} used TESS data to characterise 2MASS J06464003+0109157, an EB consisting of two 0.6\Msun\ stars. Acton et al.\ \cite{Acton+20mn,Acton+20mn2} studied NGTS J214358.5$-$380102, an M-dwarf EB with an unusually large period (7.62\,d) and eccentricity (0.323), and NGTS J0930$-$18, an EB containing a 0.58\Msun\ and a 0.08\Msun\ star. Swayne et al.\ \cite{Swayne+20mn} used the TESS data of 1SWASP J011351.29+314909.7 (Fig.\,\ref{fig:normal}) to rule out a previous suggestion that the M-dwarf secondary component was unexpectedly hot.

Mercury–manganese (HgMn) stars are late-B stars with slow rotation, no strong magnetic field, and overabundances of heavy elements. They preferentially occur in close binaries due to tidal effects slowing rotation. Paunzen et al. \cite{Paunzen+21mn} used TESS data to find that V680 Mon is an eclipsing heartbeat binary whose primary component is a HgMn star. A further seven HgMn EBs have been found by Kochukhov et al.\ \cite{Kochukhov+21mn,Kochukhov+21mn2}.

\subsection{Pulsators}                                

\begin{figure}[t]
\includegraphics[width=13.5cm]{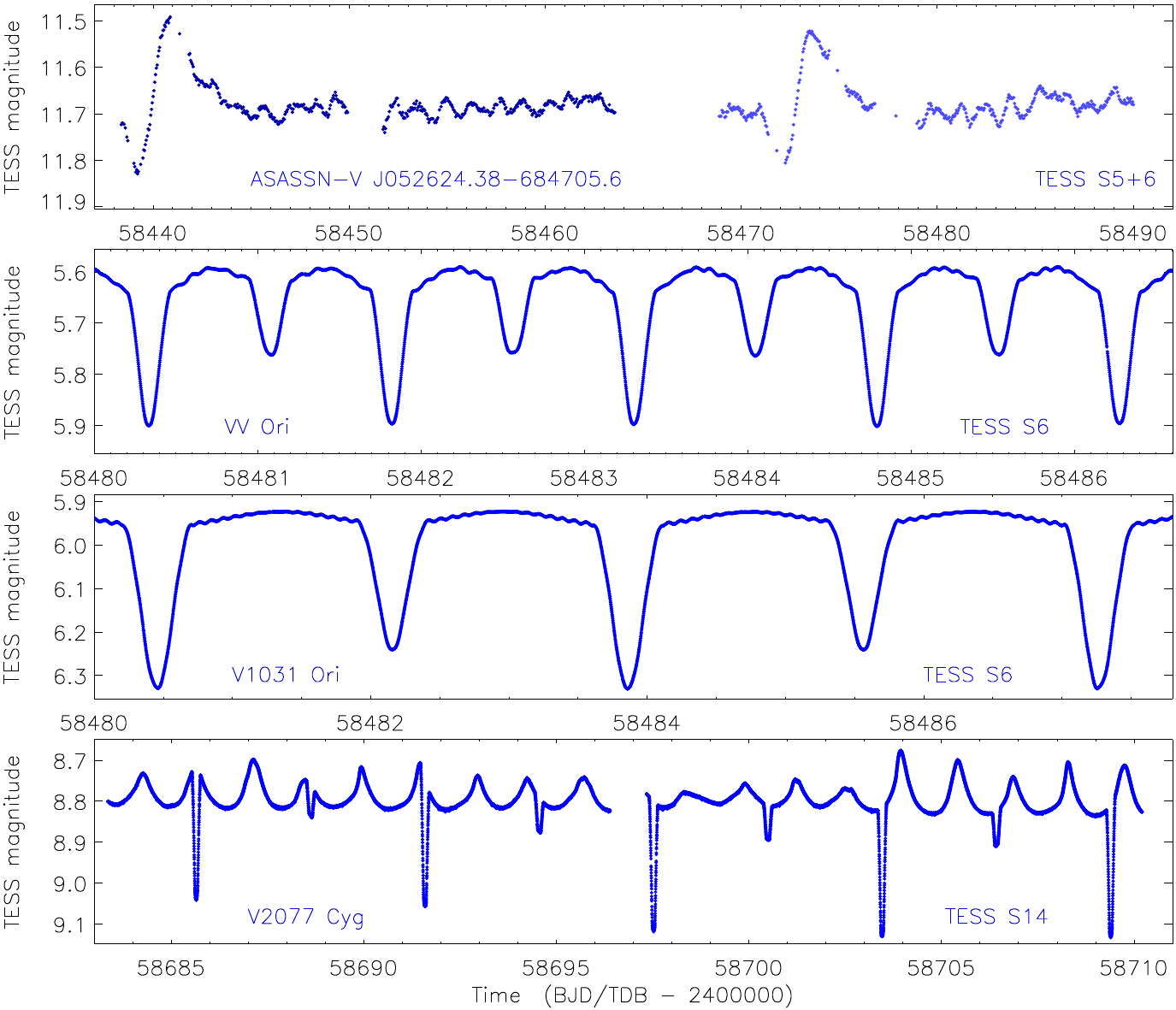}
\caption{Example TESS light curves of binary systems containing 
pulsating stars. The names and sectors are labelled. \label{fig:puls}}
\end{figure}

The large number of targets observed by TESS means many new pulsating stars in binaries have been discovered. The following discussion of pulsation types moves approximately from high-mass to low-mass.

Ko{\l}aczek-Szyma\'nski et al.\ \cite{Kolaczek+21aa} searched for high-mass heartbeat stars in data from TESS sectors 1--16. They found 20 objects, of which seven show tidally-excited oscillations. ASASSN-V J052624.38-684705.6 (Jayasinghe et al.\ \cite{Jayasinghe+19mn}) was found using ASAS-SN data and studied in detail using TESS data; its heartbeat signature is an extraordinary 40\% amplitude. The light curves from TESS sector 5 and 6 are shown in Fig.\,\ref{fig:puls}. A set of high-mass stars showing stochastic oscillations has been identified by Southworth \& Bowman (in prep.).

$\beta$\,Cephei stars are early-B dwarfs (masses approximately 8--25\Msun) which pulsate in low-radial-order p- and g-modes \cite{LeshAizenman78araa}. They have short pulsation periods ($\sim$0.1--1\,d) and pulsation amplitudes up to 0.3\,mag. Until recently none had been found in EBs suitable for precise mass and radius determination, precluding the possibility of combining binarity and asteroseismology in the study of high-mass stars. However, the huge archive of TESS light curves mean many examples are now known. Southworth et al.\ \cite{Me+20mn} found the first: V453\,Cyg is a 14\Msun\ + 11\Msun\ EB which shows total eclipses and tidally-perturbed $\beta$\,Cephei pulsations. Line-profile variations suggest the pulsations arise from the primary star, making it the first $\beta$\,Cephei star with an accurate and precise mass measurement. The system also shows apsidal motion so may become a cornerstone of massive-star evolution; new spectroscopic observations are being obtained in pursuit of this goal. A second system, CW\,Cep, has been detected by Lee \& Hong \cite{LeeHong21aj}. A third system, VV\,Ori (Southworth et al.\ \cite{Me++21mn}), shows a rich pulsation spectrum (51 significant frequencies found in one sector of TESS data) with tidal perturbations. VV\,Ori used to be totally-eclipsing but now shows partial eclipses -- this is likely due to gravitational interactions with a third body on a wider orbit \cite{Me++21mn}. The TESS data of VV\,Ori are shown in Fig.\,\ref{fig:puls}; a comparison with the OAO-2 data in Fig.\,\ref{fig:vvori} shows the difference in scatter as well as the disappearance of the total eclipse. The semi-detached binary V\,Pup shows $\beta$\,Cephei pulsations (Budding et al.\ \cite{Budding+21mn}), and a further 12 $\beta$\,Cephei EBs have been discovered by Southworth \& Bowman (in prep.).

SPB stars show multi-periodic g-mode pulsations with periods of 1--4\,d and are valuable for probing the internal structure of massive stars (Waelkens \cite{Waelkens91aa}). They are slowly-rotating dwarfs of spectral type late-B and early-A hence should be fairly common in EBs, but few are known. The first with precise measurements of its mass and radius was V539\,Ara (Clausen \cite{Clausen96aa}). The secondary component of VV\,Ori might show SPB pulsations \cite{Me++21mn}. Stassun et al.\ \cite{Stassun+21apj} found several important results for HD\,149834: it is an EB with a period of 4.60\,d; it is a member of the young open cluster NGC\,6193; the primary component is a hybrid that shows both $\beta$\,Cephei and SPB pulsations; and it shows magnetic activity in the form of a flare and possible rotational modulation.

Another extremely unusual phenomenon is what has been referred to as single-sided or tidally-trapped pulsations, where the pulsation amplitudes vary across the surface of a star. This has been detected in HD\,74423, an ellipsoidal variable with a $\delta$\,Scuti pulsator, by Handler et al.\ \cite{Handler+20natas}. Its light curve was interpreted as being due to the pulsation axis of a dipole-pulsating star being aligned with the tidal deformation axis. CO\,Cam is an Am star, showing $\delta$\,Scuti oscillations, in a binary system with a G star (Kurtz et al.\ \cite{Kurtz+20mn}). The oscillations have a larger amplitude in the region of the Am star that is pointing towards the G star, although both stars are well inside their Roche lobes so are relatively undistorted. TIC 63328020 was found by Rappaport et al.\ \cite{Rappaport+21mn} to show tidally tilted pulsations of the $\delta$\,Scuti type, making it the third object in this class but with very different properties. Both stars are close to filling their Roche lobes, and their masses (2.5\Msun\ and 1.1\Msun) and radii (3.0\Rsun\ and 2.0\Rsun) can only be matched by stellar theory if the binary has undergone extensive mass transfer. V1031\,Ori (Lee \cite{Lee21pasj}) is a triple system showing eclipses and $\delta$\,Scuti pulsations on one side of the primary star (Fig.\,\ref{fig:puls}), making it the fourth object known in this class of pulsator.

Perhaps the most common pulsation mechanism detected within binary systems is $\delta$\,Scuti, partly because the short pulsation periods are relatively easy to find in a single night of ground-based photometry. TESS can nevertheless improve on this by providing light curves with a higher precision and without diurnal interruptions. $\delta$\,Scuti oscillations have been found in the eccentric and totally-eclipsing binary AI\,Hya. Lee et al.\ \cite{Lee+20pasj} measured 14 frequencies, although none coincide with the single frequency measured by J{\o}rgensen \& Gr{\o}nbech \cite{JorgensenGronbech78aa} from ground-based photometry. Lee et al.\ \cite{Lee+20pasj} also determined the masses and radii of the stars and detected slow apsidal motion. The triple system IM\,Per has been found to show $\delta$\,Scuti, $\gamma$\,Dor and tidally-perturbed oscillations in TESS data (Lee et al.\ \cite{Lee+21aj}), as has RR\,Lyn (Southworth, \cite{Me21obs6}). An important recent result for $\delta$\,Scuti pulsations is the discovery of regular sequences of oscillations in young stars which enable clear mode identification \cite{Bedding+20nat}. This is important for enabling the use of asteroseismology to determine the properties of $\delta$\,Scuti stars, and scaling relations have been established using TESS data by Barcel\'o Forteza et al.\ \cite{Barceloforteza+20aa} and Hasanzadeh et al.\ \cite{Hasanzadeh++21mn}.

EBs with $\gamma$\,Dor pulsations have been found in TESS data. Most also show $\delta$\,Scuti and tidally-perturbed pulsations (e.g.\ IM\,Per and RR\,Lyn from the last paragraph). A set of five new discoveries is presented by Southworth \& Van Reeth (in prep.) including V2077\,Cyg (Fig.\,\ref{fig:puls}) which is strikingly reminiscent of KIC 11285625 (Fig.\,\ref{fig:kepler}).

\subsection{Multi-eclipsers}                          
\label{sec:tess:multi}

Multiple systems showing multiple types of eclipse are rare and require extensive data to study in detail. In comparison to \textit{Kepler}, TESS provides far shorter datasets but for vastly more point sources. TESS is thus a rich source of multiple-eclipsing systems, although additional data such as from SuperWASP \cite{Pollacco+06pasp} can be critical in lengthening the time coverage of the data for specific objects.

Borkovits et al.\ \cite{Borkovits+20mn2} presented TIC 167692429 and TIC 220397947, both of which are hierarchical triple systems where the third body causes eclipse timing variations. The former system also displays changes in eclipse depth due to precession of the orbital plane caused by a misalignment between the inner and outer orbits. In contrast, TIC 278825952 (Mitnyan et al.\ \cite{Mitnyan+20mn}) is a triple system where both orbits are circular and coplanar. Borkovits et al.\ \cite{Borkovits+20mn} analysed TIC 209409435, which shows additional eclipses as well as eclipse timing variations due to a third body. They obtained a complete description of the system using the TESS and ground-based photometry, including constraints from the {\sc parsec} theoretical stellar evolutionary models as a substitute for RV measurements.

Rowden et al.\ \cite{Rowden+20aj} found that TIC 278956474 is a quadruple system containing two eclipsing binaries (periods 5.49 and 5.67\,d) orbiting each other with a period of 858\,d. From the eclipses, two RVs and eclipse time variations, they determined the properties of the four stars. The stars have masses from 0.45 to 1.3\Msun\ and radii from 0.48 to 1.6\Rsun. Such systems have the potential to be a strong test of stellar models because the four stars have different masses and radii but the same age and chemical composition, but TIC 278956474 itself is not a good candidate for this because RVs are difficult to measure for all but the brightest of the components. A better possibility is BG\,Ind (Borkovits et al.\ \cite{Borkovits+21mn}) which contains two EBs (periods 1.46 and 0.52\,d) with an outer orbit of 721\,d. The two components of the brighter EB are comparatively straightforward for RV measurement and their properties could be established to high precision. Even better is TIC 454140642 (Kostov et al.\ \cite{Kostov+21apj}) which shows two sets of eclipses on periods of 13.6\,d and 10.4\,d, eclipse timing variations due to an outer orbit of period 432\,d, and four sets of lines in high-resolution spectra. The four stars are very similar, having masses 1.1--1.2\Msun\ and radii 1.1--1.3\Rsun. BU\,CMi has also been found to show two sets of eclipses of similar depth (Volkov et al.\ \cite{Volkov++21}).

The most spectacular multi-eclipser, though, is TIC 168789840 (Powell et al.\ \cite{Powell+21aj}). This consists of three EBs, all similar to each other, with periods of 1.57, 8.21 and 1.31\,d. The first two EBs orbit each other every 3.7\,yr and the third EB (spatially resolved at a separation of 0.42$^{\prime\prime}$) is on a wider orbit of roughly 2000\,yr. The primary component of each EB is 1.2--1.3\Msun\ and 1.5--1.7\Rsun, and the secondary components are all approximately 0.6\Msun\ and 0.6\Rsun. The properties of this system may in future help to understand the process of binary star formation as well as the dynamical evolution of such systems.

\subsection{The connection between binaries and planets} 

As mentioned in Section\,\ref{sec:kepler:planet}, a total of 14 transiting circumbinary planets are known. Whilst 12 were found using {\it Kepler} data, two have been identified from TESS observations. TOI-1338 (Kostov et al.\ \cite{Kostov+20aj}) was a known EB with a 14.6\,d orbit, originally detected as a planet candidate in the SuperWASP survey but found to be low-mass EB instead. It was observed for almost a year using TESS, as it is in the southern continuous viewing zone, and in this time three transits were seen of the same depth but very different duration. The masses and radii of all three components (two stars and one planet) were determined with good precision from these data and RVs of the bright primary star. One more transiting circumbinary planet has been found in TESS data: TIC 172900988 is clearly detected but there are several alternative solutions for its mass and orbital period so additional observations are needed for a precise characterisation of the system (Kostov et al.\ \cite{Kostov+21}).

Surveys for transiting planets typically throw up a lot of false positives -- objects that show a transit of suitable depth (less than about 3\%) and duration but are not actually planetary systems. A significant fraction of these are EBs containing a low-mass star, because late-M stars can have the same radii as gas-giant planets (e.g.\ \cite{Me11mn}). Lendl et al.\ \cite{Lendl+20mn} identified TOI-222 as a possible transiting planet showing a single transit in TESS observations, but found it to be an EB with a 33.9\,d period. The component stars have masses 1.1\Msun\ and 0.23\Msun\ and radii 1.0\Rsun\ and 0.18\Rsun, so the secondary has a radius consistent with that of an inflated giant planet. A very similar result was found for TIC 231005575 by Gill et al.\ \cite{Gill+20mn}: it is a 61.8\,d EB with a G-type primary (1.0\Msun\ and 1.0\Rsun) and a low-mass secondary (0.13\Msun, 0.15\Rsun) masquerading as a planet.

\subsection{Semi-detached binaries}                   

\begin{figure}[t]
\includegraphics[width=13.5cm]{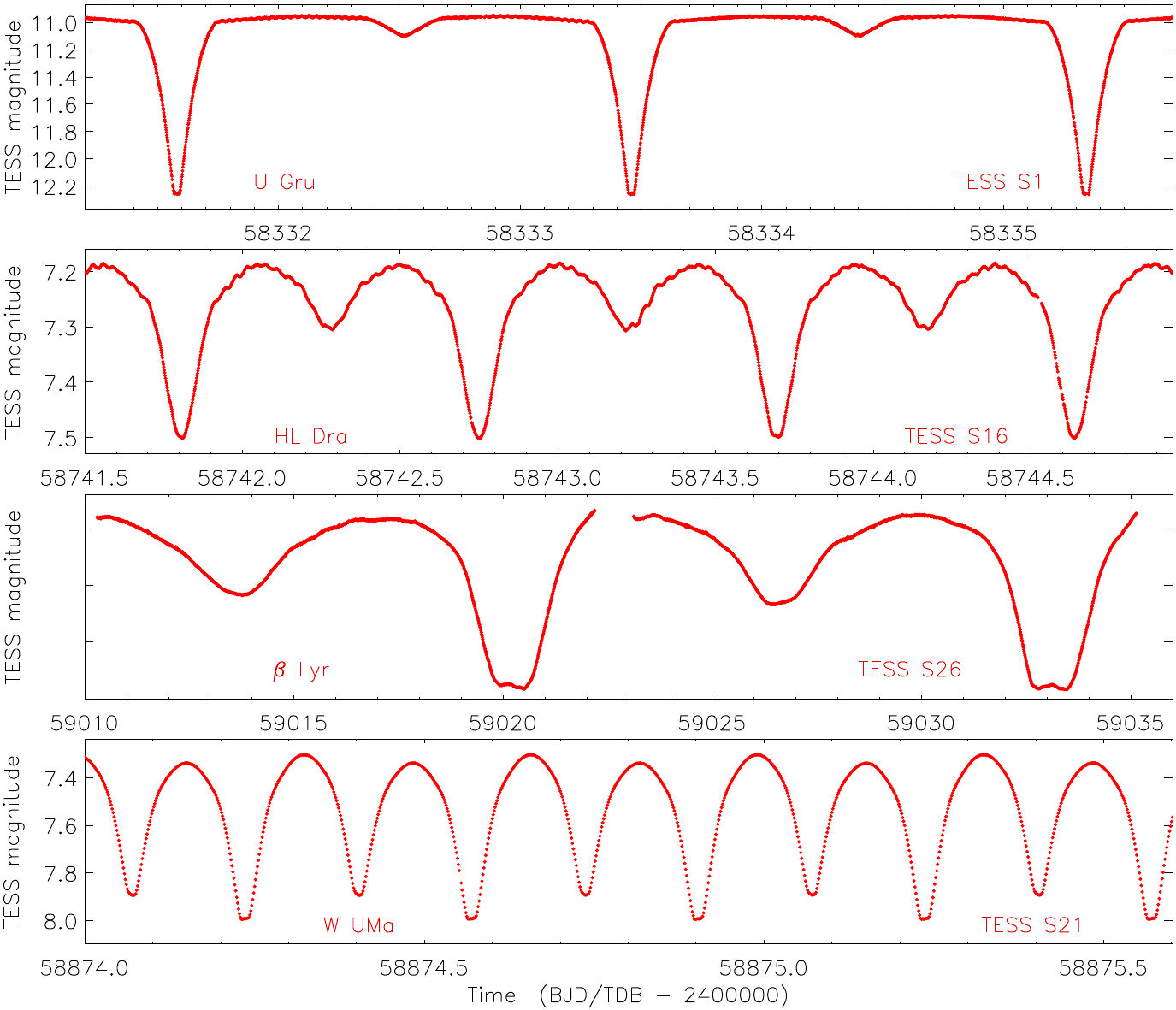}
\caption{Example TESS light curves of semi-detached and contact systems. The y-axis for
$\beta$\,Lyr is not labelled as the eclipse depths in the TESS data are incorrect. \label{fig:sd+c}}
\end{figure}

Relatively little work has been done on semi-detached binaries so far with TESS. The light curve of the prototype eclipsing system $\beta$\,Lyr is shown in Fig.\,\ref{fig:sd+c} as an example.

Miller et al.\ \cite{Miller+21aj} characterised 2MASS J17091769+3127589 using the TESS light curves and RVs from the APOGEE project. They found it to consist of a 1.5\Msun\ and 2.6\Rsun\ subgiant primary and a 0.26\Msun\ and 4.0\Rsun\ giant secondary. Binary evolution models from the {\sc mesa} code were able to fit these properties for a system with initial masses of 1.2\Msun\ and 1.1\Msun\ and period 3.43\,d, so this system has undergone an extreme mass ratio reversal via case B mass transfer.

Bowman et al.\ \cite{Bowman+19apj} found p-mode oscillations in the semi-detached EB U\,Gru (Fig.\,\ref{fig:sd+c}), some untouched and some perturbed by tidal effects. They identified three possible sources of the rich frequency spectrum of this system, all of which require tidal effects; high-resolution spectroscopy is needed for further analysis. HL\,Dra has also been found to be a semi-detached system (Fig.\,\ref{fig:sd+c}) with a pulsating primary component (Shi et al.\ \cite{Shi+21mn}). A total of 252 pulsation frequencies were found, of which some are tidally perturbed.

\subsection{Contact binaries}                         

Contact binaries are generally not well-suited to space-based photometry because the low sampling rate (e.g.\ 1765\,s for \textit{Kepler} long cadence and 1800\,s for full-frame images early in the TESS mission) and their short orbital periods (as little as 0.2\,d) mean their light variations can be severely undersampled. However, a small fraction have been observed at a higher cadence (59\,s for \textit{Kepler}, 120\,s for TESS) and their spectral types are often within the range of stars prioritised in searches for transiting planets. Part of the TESS light curve of the prototype contact EB W\,UMa, taken in short cadence, is shown in Fig.\,\ref{fig:sd+c} and can be contrasted with the \textit{Hipparcos} observations in Fig.\,\ref{fig:hip}.

Xia et al.\ \cite{Xia+21pasp} studied ASAS J124343+1531.7 and LINEAR 2323566 using data from the TESS full-frame images. NSVS 5029961 was also observed in the TESS full-frame images and these data were used by \citet{Zheng+21mn} to determine the properties and evolutionary status of the components. The CoBiToM project (Gazeas et al.\ \cite{Gazeas+21mn}) aims to study the merger of contact binaries using data from TESS among other ground- and space-based telescopes.

\subsection{Evolved binaries}                         

This section is dedicated to binaries containing evolved objects such as white dwarfs. These are typically faint so not well suited to study with a small telescope optimised for bright stars, but the huge sky coverage of TESS means that a good selection of the brighter examples have been observed.

Wang et al.\ \cite{Wang++20apj} used the TESS database to identify two new EL\,CVn systems, which are short-period EBs containing a normal A star and a hot low-mass pre-white-dwarf \cite{Maxted+11mn,Maxted+14mn}. In both cases $\delta$\,Scuti pulsations were detected from the A star and higher-frequency pulsations were also found which might arise from the secondary component. Sahoo et al.\ \cite{Sahoo+20mn} and Baran et al.\ \cite{Baran+21mn,Baran+21mn2} have searched for and found several new HW\,Vir systems -- short-period EBs containing a subdwarf B star and a late-type dwarf.

A modest number of bright CVs are known and some have useful observations from TESS. The eclipsing dwarf nova EM\,Cyg was observed in one sector by TESS, and an outburst with quasi-periodic oscillations was seen (Liu et al.\ \cite{Liu+21mn}). Schaefer \cite{Schaefer21rnaas} determined orbital periods for several classical novae from TESS observations.

Some CVs contain a magnetic white dwarf. Depending on the field strength the magnetic field may truncate (in intermediate polars) or entirely disrupt (in polars) the accretion disc. TESS observations of the intermediate polar TX\,Col have been used to determine its orbital, spin and beat periods and to detect quasi-periodic oscillations \cite{Rawat+21apj,Littlefield+21aj}. A revision of the orbital and spin periods of the polar CD\,Ind based on TESS data led to a new interpretation of its accretion (Littlefield et al.\ \cite{Littlefield+19apj}),

AM\,CVn systems are the more evolved friends of CVs, containing two white dwarfs where the larger and less massive one fills its Roche lobe and undergoes mass loss. The outbursts and superoutbursts of these objects remain poorly understood due to the diversity of behaviour as a function of orbital period (Duffy et al.\ \cite{Duffy+21mn}). TESS has seen outbursts of several AM\,CVn systems and these observations are leading to an improved understanding of these extreme objects \cite{Duffy+21mn,Pichardo+21mn}.

\begin{figure}[t]
\includegraphics[width=13.5cm]{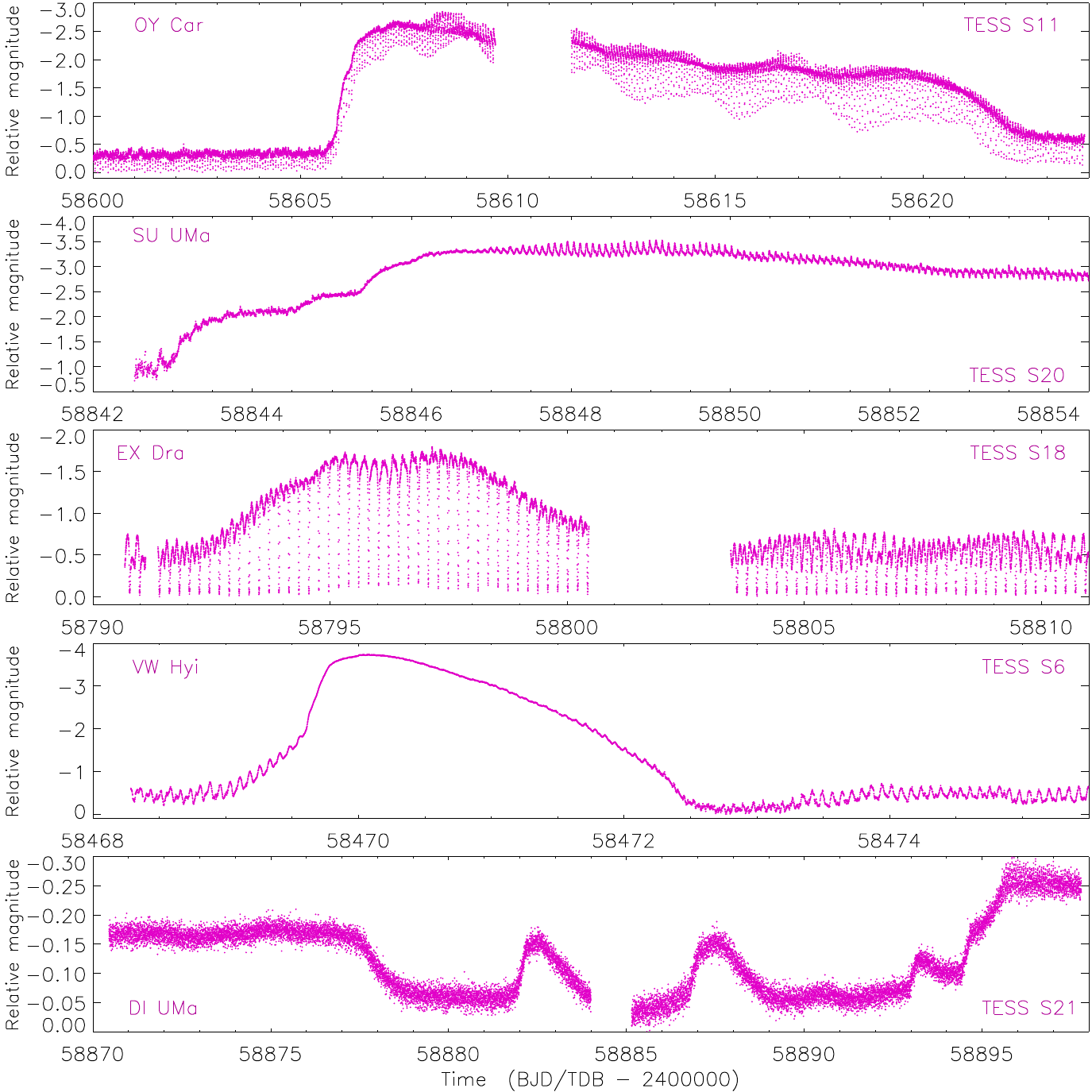}
\caption{Example TESS light curves of CVs. The magnitude scales are not reliable due to the faintness of these objects for TESS. \label{fig:cv}}
\end{figure}

Fig.\,\ref{fig:cv} shows a set of TESS light curves of CVs, many of them well-known systems, picked out by the author to illustrate the variety of behaviours of these objects and the quality of the TESS data for them. OY\,Car is an eclipsing CV caught in superoutburst in TESS sector 11 -- the eclipse depth can be seen to change dramatically throughout the superoutburst. SU\,UMa has again been caught in superoutburst and exhibits a particularly lovely set of superhumps in the second half of the plot. Also of interest are the several plateau phases during the rise to peak brightness. EX\,Dra is a long-period CV (5.0\,hr) which shows eclipses of extraordinary depth during outburst, because the large size of the secondary component means it fully blocks the white dwarf and accretion disc during eclipse. VW\,Hyi has strong orbital variability which disappears during outburst and reappears afterwards. DI UMa is a little-known dwarf nova that appears to show a wide range of variability states during TESS sector 21; one possible interpretation is that it is a novalike system and the magnitude scale of the TESS data has been compressed. Detailed analyses of these objects would need a careful re-reduction of the TESS data using methods optimised for faint stars.


\section{Discussion}

This review has traced the development of space-based time-series photometry of binary systems from OAO-2, the first space telescope, to TESS, the most recent one. This development has passed through three main phases due to the available technology and the science cases of particular projects.

The first phase covers the early missions such as OAO-2 (launched in 1968), ANS (1974), Voyagers 1 and 2 (1977) and IUE (1978). These were primarily aimed at providing UV observations because this wavelength range is not accessible from below Earth's atmosphere. The throughput was low, due to the generally small size of the telescopes and the sensitivity of the UV light detectors then available. This meant that only UV-bright stars could be studied, which dictated hot stars with bright apparent magnitudes, and thus the EBs observed were high-mass and usually short-period. Interesting results were obtained for a range of science areas, including physical properties, limb and gravity darkening, starspots, mass transfer and circumstellar matter.

In the second phase are several satellites which observed at visual wavelengths and, in some cases, were not necessarily intended to provide time-series photometry. These missions did not cover a wavelength range inaccessible from Earth's surface, but instead provided data of greater quality and quantity than normally achievable from the ground. Their location in space allowed them to observe without interruption for daylight or bad weather, and without the scintillation noise that afflicts ground-based telescopes. The \textit{Hipparcos} mission had an overriding goal of obtaining precise measurements of the \emph{position}, not \emph{brightness}, of celestial objects, but nevertheless returned light curves of typically 100 or more datapoints for many bright stars. The \textit{Hipparcos} data led to the discovery of a large number of bright EBs, some of which have subsequently been studied in detail. The WIRE satellite was launched to obtain IR imaging of galaxies but the failure of the main mission opened up the possibility to use the star-tracker telescope to obtain light curves of stars of naked-eye brightness. WIRE obtained the first high-precision and well-sampled light curves of EBs from space. The MOST satellite was designed to obtain high-precision and long-duration photometry of individual objects; it was a huge success for asteroseismology but did not observe many binary stars. The BRITE-Constellation of satellites are a development of this theme.

We are now in phase 3: space telescopes dedicated to obtaining high-precision photometry for many stars simultaneously for asteroseismology and searching for planetary transits. \textit{CoRoT} was the first of these missions and obtained light curves of several thousand EBs among the over 160\,000 stars observed. Whilst a significant step forward, \textit{CoRoT} was a small telescope observing relatively faint stars so significant effort was required to follow up any interesting objects it found. This was particularly the case for the transiting planets it discovered \cite{Deleuil+18aa}. Then came \textit{Kepler}, whose archive of photometry remains the highest quality available. TESS is the current figurehead of phase 3: its data are significantly inferior to \textit{Kepler} in both precision and duration, but its huge sky coverage has led to major advances in many fields of binary star research.

A feature of the phase-3 space missions is their concentration on finding transiting planets, which leads to a requirement of high-precision photometry obtained over long timescales. EBs are typically not part of the main science case of these missions but hugely benefit from the data quality. In searches for transiting planets one will always find EBs because their eclipse depths are usually much greater. These EBs must then be weeded out from the transit search but can themselves provide important new results: one person's rubbish is another person's objects of interest. We thus find ourselves in the position where people studying planets reject EBs as they are not what they are looking for; people working on EBs often find that any pulsations get in the way and pass them on to asteroseismologists; and asteroseismologists typically remove eclipses and planetary transits before analysing the pulsations. This is a virtuous circle where there is something interesting for all three disciplines to work with. The implication is that people studying planets, binaries, or pulsations should work closely together, and this is indeed becoming more common.

\subsection{Current status}

Important scientific results have been obtained from the previous and current sets of space missions in many areas of binary-star science. This includes studies of high-mass stars in EBs; low-mass stars where the radius discrepancy remains unexplained; semi-detached and contact binaries which evolve through phases of mass transfer and mass loss; binaries containing a wide variety of pulsating stars ($\delta$\,Scuti, $\gamma$\,Dor, SPB, $\beta$\,Cephei, stochastic); binaries with giant stars (which often display solar-like oscillations); triple systems showing extra eclipses and variations in the times and durations of eclipses; chemically-peculiar stars in EBs; doubly (and even triply) eclipsing systems; planets orbiting EBs; self-lensing binaries; binaries with subdwarf B stars; CVs (including eclipses, outbursts, superoutbursts, novae, magnetic systems); and X-ray binaries. The deluge of data from TESS means there is still a lot more to be studied.

\begin{figure}[t]
\includegraphics[width=13.5cm]{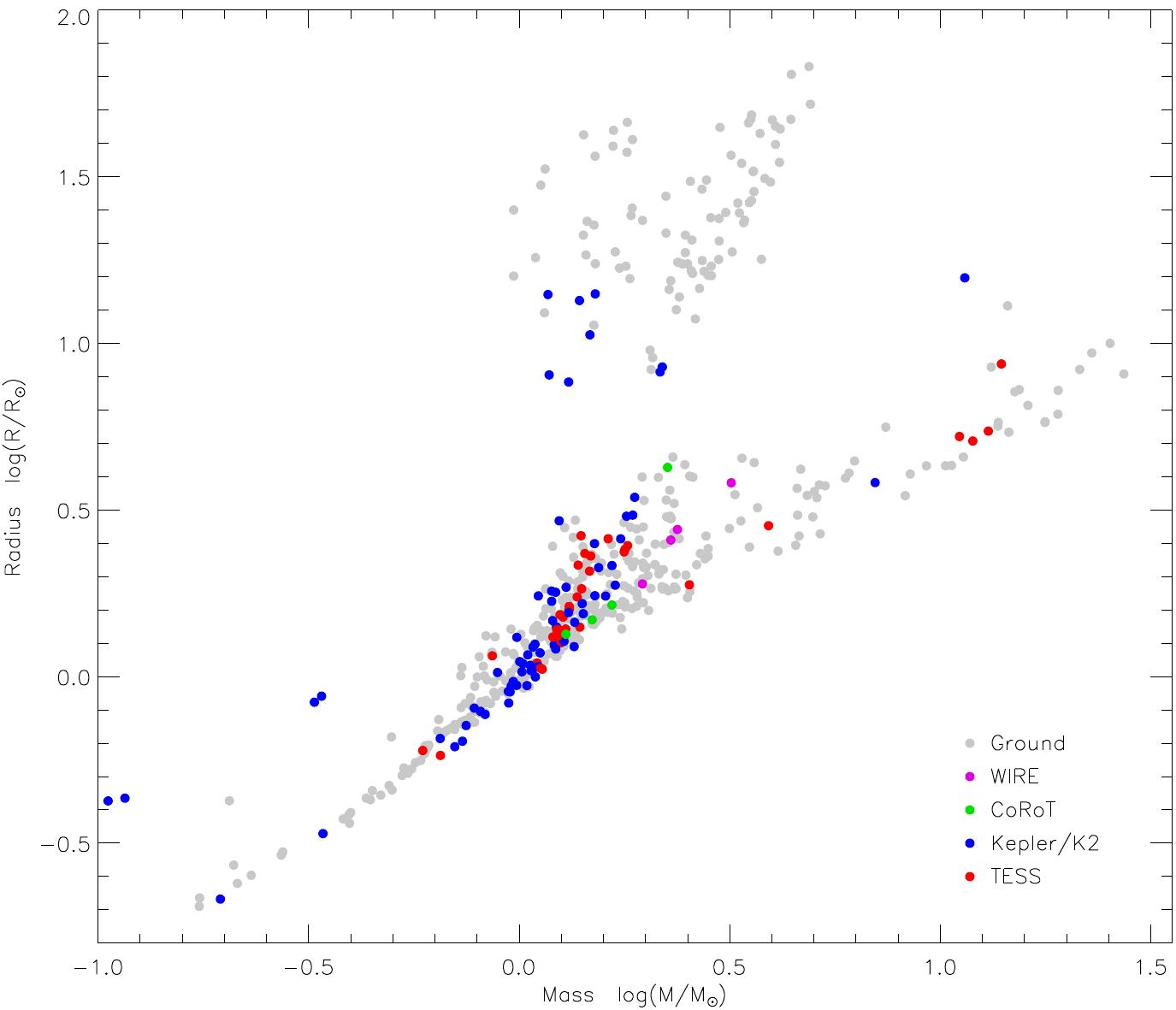}
\caption{Mass-radius diagram for EBs in the DEBCat catalogue \cite{Me15debcat}. Results from space-based telescopes are colour-coded
according to source. Errorbars are not plotted as they are almost all smaller than the point size. \label{fig:mr}}
\end{figure}

\begin{figure}[t]
\includegraphics[width=13.5cm]{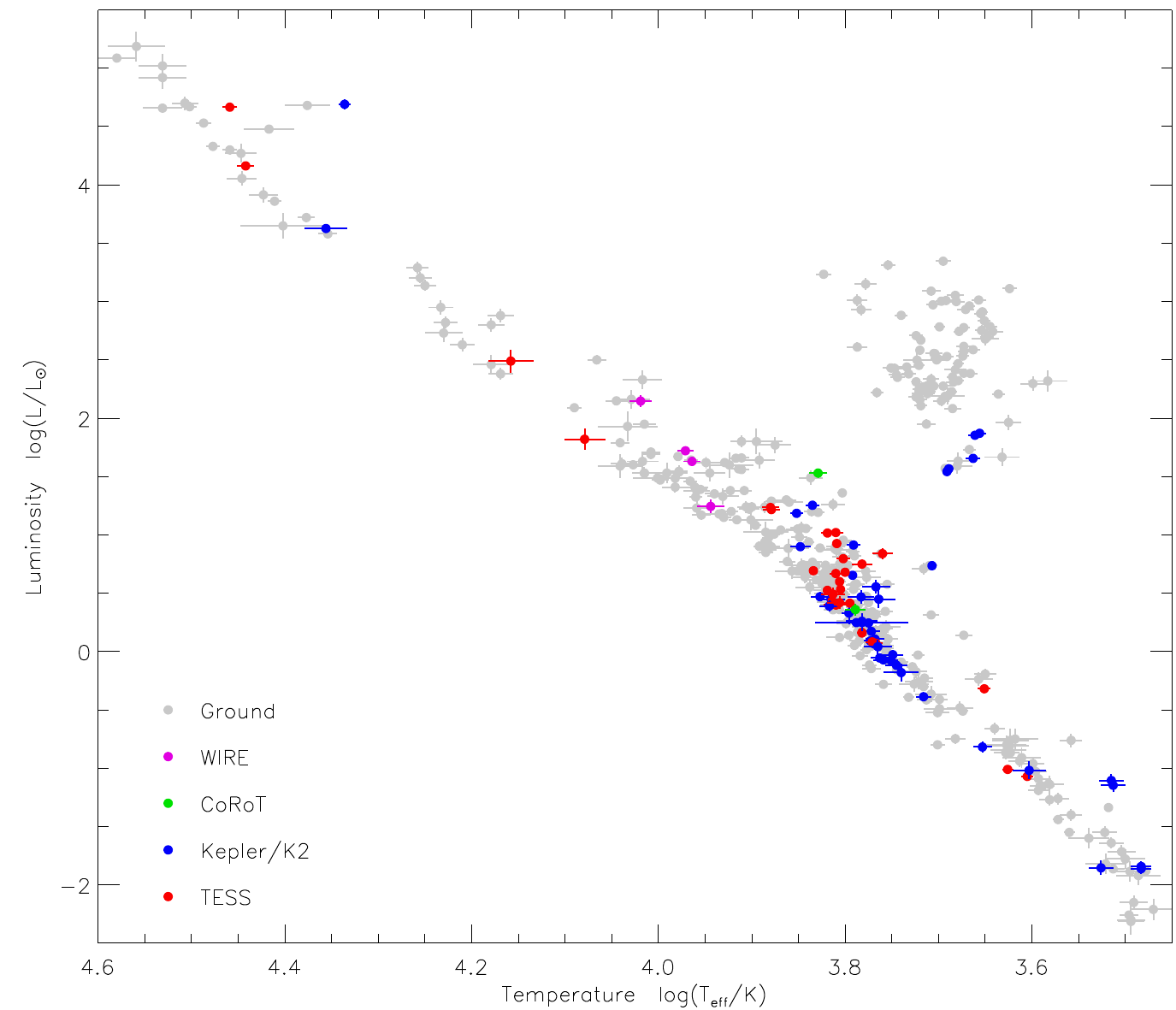}
\caption{Hertzsprung-Russell diagram for EBs in the DEBCat catalogue \cite{Me15debcat}.
Results from space-based telescopes are colour-coded according to source. \label{fig:hrd}}
\end{figure}

One way to visualise the impact of space missions is by plotting the properties of EBs in mass-radius (Fig.\,\ref{fig:mr}) and Hertzsprung-Russell (Fig.\,\ref{fig:hrd}) diagrams. The stars plotted here are those with mass and radius measurements to 2\% or better collected in DEBCat (the Detached Eclipsing Binary Catalogue\footnote{\texttt{https://www.astro.keele.ac.uk/jkt/debcat/}}; Southworth \cite{Me15debcat}). Light grey points indicate properties established from ground-based data whilst coloured points show those based on light curves from space missions. Most points are still from ground-based observations, but a significant number are from \textit{Kepler} and TESS. \textit{Kepler} dominates TESS at lower masses and larger radii, as \textit{Kepler} studied fainter stars on average and also observed for much longer so was well suited to the analysis of EBs containing giant stars. However, TESS is already out-performing \textit{Kepler} at higher masses due to its much larger sky coverage.

\subsection{Future missions}

Recent work on \textit{CoRoT}, \textit{Kepler} and TESS has shown how much binary stars can contribute to astrophysics. However, they are not as flashy, or as saleable to the general public, as extrasolar planets. Future missions will therefore likely follow a similar template of searches for transiting planets on which binary star science can hitch a lift. This is not a bad thing, as planet searches require a higher data quality than most binary star studies.

An example is CHEOPS (CHaracterising ExOPlanets Satellite), launched by ESA in December 2019 and currently operating successfully in a sun-synchronous polar orbit around Earth. Its primary science goal is detect transits of planets already known from RV measurements (Benz et al.\ \cite{Benz+21exa}), which has some overlap with TESS. Its concentration on single objects, however, makes it more flexible than TESS. The first results of a programme to study low-mass EBs have now been published (Swayne et al.\ \cite{Swayne+21mn}).

An important future contributor to the rapidly-growing archives of space-based photometry is the \textit{Gaia} satellite \cite{Gaia16aa}, which is currently obtaining up to 200 epochs of high-precision photometry for over one billion stars. Once released these data will allow a search for EBs complete to much fainter magnitudes than previously achieved. \textit{Gaia} will obtain far fewer observations per star than \textit{Kepler} or TESS so will be less effective at finding EBs that spend only a small fraction of their time in eclipse. A unique advantage of \textit{Gaia} is that it will also obtain high-resolution spectroscopy simultaneously with the photometry and astrometry, so will yield sufficient data to measure the masses and radii of a significant fraction of the short-period EBs it observes.

Further ahead lies the flagship mission for ESA in planetary and stellar science: PLATO (PLAnetary Transits and Oscillations of stars) is currently scheduled for launch in 2026. It will take a multi-telescope approach to obtain high-precision photometry of a large number of bright stars for several years in order to find rocky planets with short and intermediate orbital periods around Sun-like stars (Rauer et al.\ \cite{Rauer+14exa}). PLATO is expected to provide photometry significantly better than achieved by \textit{Kepler} but for a shorter duration, and cover a sky area significantly larger than \textit{Kepler} but still much smaller than TESS. Its concentration on F, G and K stars means it may detect a large number of solar-type and low-mass EBs with a wide range of orbital periods, a set of objects which will be very important in calibrating theoretical models of stars in the lower main sequence (i.e.\ planet host stars). PLATO is also explicitly depending on asteroseismology of solar-type stars to return precise stellar properties (mass, radius and age) of the host stars of the transiting planets it will find. Once again, EBs provide model-independent quantities that can be used to calibrate asteroseismology (e.g.\ \cite{Themessl+18mn,Benbakoura+21aa}). The trickiest measurable here is stellar age -- this is not directly accessible in a typical EB but can be obtained if the EB is a member of a stellar cluster (e.g.\ \cite{Me++04mn,Brogaard+11aa}).

It should also be remembered that there are many ground-based large photometric surveys which yield data of lower quality but larger quantity than the space missions discussed in this review. Including these observations can greatly extend the time coverage of a dataset compared to using the space-based data alone, and this approach is already being successfully used for work on multiply-eclipsing systems (see Sections \ref{sec:kepler:multi} and \ref{sec:tess:multi}). Past surveys were primarily aimed at finding transiting planets, e.g.\ TrES \cite{Alonso+04apj}, HAT and HATSouth \cite{Bakos+02pasp,Bakos+13pasp}, SuperWASP \cite{Pollacco+06pasp} and KELT \cite{Pepper+07pasp}. Current and future surveys have become even more ambitious in sky coverage, aiming to monitor the entire observable sky simultaneously whenever conditions allow (e.g. MASCARA \cite{Talens+17aa}, HATpi\footnote{\texttt{https://hatpi.org/}} and Evryscope \cite{Law+15pasp}). Further ahead, the Rubin Observatory will find large numbers of faint EBs (Geller et al.\ \cite{Geller+21apj}).

\subsection{Future work}

There are many areas where our understanding should be improved. Here is a (partial) list:

\textbf{Low-mass stars.} The radius discrepancy has not yet been solved. What would help would be a large sample of low-mass stars in EBs with precise measurements of their mass, radius, \Teff, chemical composition, rotation rate, magnetic activity and age. A multivariate analysis of these results then might point to the source(s) of the problem. The age is likely to be the trickiest property to measure, because the evolutionary timescales are extremely long. However, age could be inferred for EBs via kinematics or membership of a stellar cluster. Another method to constrain the age of low-mass stars in EBs is to study binaries containing an F- or G-type dwarf primary star, for which the age is measurable from its mass and radius, and an M-dwarf secondary star.

\textbf{Solar-type stars.} The theoretical understanding of solar-type stars is crucial (e.g.\ for inferring the properties of planets via their host stars) but we remain unable to properly understand the Sun. A large sample of solar-type stars in EBs would help to calibrate some of the fudge factors in theoretical stellar models. Again, measurements of mass, radius, \Teff, chemical composition, rotation, magnetism and age would be useful. Many such measurements currently exist but need to be more precise. This work is of particular interest because the chemical abundances of the Sun are currently controversial \cite{Asplund+09araa,Asplund++20xxx}.

\textbf{Calibrating theoretical models.} There are two main requirements for EBs for them to be a useful probe of theoretical models. First, the two components should have significantly different properties to make it more difficult to find a theoretical model that fits both simultaneously. Second, the two components should have a similar mass so any problems with theoretical models can be assigned to a particular mass range. These two requirements are in obvious conflict. They can be reconciled by measuring the properties of ``benchmark'' EBs containing similar stars, to a very high precision (approaching 0.1\% in mass and radius), so their properties are similar but still differ by much more than the errorbars. This is difficult to do, but is possible with space-based light curves of suitable systems \cite{Maxted+20mn}.

\textbf{A set of benchmark EBs.} Benchmark EBs -- ones where radius can be measured to 0.1\% -- are not easy to find. It is best if they are totally-eclipsing, because the times of the four contact points in the eclipse map directly into the radii of the stars (Kopal \cite{Kopal59book}) and thus allow precise measurement. They should also be well-detached for two reasons. First, stars that are well-separated experience negligible tidal effects so have evolved as single stars. Second, proximity effects (reflection effect, ellipsoidal effect, gravity darkening) are more complex to model and can make it impossible to determine radii to 0.1\% precision (e.g.\ \cite{Me20obs}). A benchmark EB must also have high-quality \Teff\ measurements (a recent approach by Miller et al.\ \cite{Miller++20mn} is promising) and spectroscopic chemical abundances \cite{Serenelli+21aarv}.

\textbf{High-mass stars.} Massive stars evolve quickly and have a large effect on their local environment as they do so. They are important in the chemical evolution of galaxies and the production of exotic objects such as supernovae and gravitational wave events. The understanding of some of the interior physics of massive stars is still limited, but would be improved by direct measurements of the properties of massive stars in EBs.

\textbf{Pulsating stars.} Thanks to \textit{Kepler} and TESS we now know many EBs containing stars with $\beta$\,Cephei, SPB, $\delta$\,Scuti and $\gamma$\,Dor pulsations. Solar-like oscillations have been found in giant stars but not dwarf stars in EBs. Pulsations provide an extra set of information on a star, and thus additional constraints that can be applied to theoretical models. $\beta$\,Cephei, SPB and $\gamma$\,Dor pulsations are most valuable because they can be sensitive to the properties deep inside stars. One complication is that pulsation modes are affected by the tidal deformation of stars so differ between single stars and those in close binaries. An important goal is to study solar-like oscillations in the components of solar-type EBs, but the extreme requirements on data quality mean this has not yet been achieved. This is needed to directly challenge the asteroseismic scaling relations for dwarf stars, as the vast majority of the known planet hosts are unevolved stars.

\textbf{Stellar clusters.} Studying EBs which are members of open or globular clusters allows more constraints to be placed on theoretical stellar models. In some cases several EBs in the same cluster are known, so the models have more constraints to match simultaneously. This has happened for clusters such as NGC\,869 \cite{Me++04mn}, NGC\,6791 \cite{Brogaard+11aa} and Ruprecht 147 \cite{Torres+20apj}. It can also be achieved with multiple systems containing more than one EB.

\textbf{Multiplicity.} The fraction of stars that are double or multiple is known to depend on mass \cite{DucheneKraus13araa} and to be obtainable from samples of EBs \cite{Shan++15apj}. Quantifying this helps in understanding the formation processes of single and binary stars \cite{MoeKratter19xxx}. The distribution of mass ratios is also an observable and theoretically interesting quantity \cite{Bate09mn}, and the eccentricity distribution can be used to test tidal theories \cite{JustesenAlbrecht21apj}.

\textbf{Evolved binaries.} Mass transfer in semi-detached and contact binaries remains poorly understood \cite{Eggleton11book,Dervisoglu+18mn} and can lead to stellar mergers \cite{Tylenda+11aa}. The common-envelope phase of binary evolution is also difficult \cite{Izzard+12iaus,Politano21aa}, and once a binary has got past these obstacles it exhibits many complex phenomena \cite{Warner95book}. The characteristics of outbursts and superoutbursts in CVs require extensive photometry -- space missions can contribute by long-term monitoring of evolved binaries and by catching the onset of outbursts (see Fig.\,\ref{fig:cv}).

\subsection{Final remarks}

Space-based photometry of binary systems can contribute to all the scientific studies listed above. In some cases (e.g.\ pulsations, population studies) it is the only feasible way to obtain the necessary data for a large number of stars. The huge database currently being created by TESS will lead to major advances for many years, and beyond that PLATO will bring a new level of photometric precision. What we need now is enough researchers to take advantage of these unprecedented opportunities.


\funding{This research received no external funding.}

\dataavailability{The data in Figs.\ \ref{fig:compare} and \ref{fig:kepler}--\ref{fig:cv} were provided by NASA and obtained from the Mikulski Archive for Space Telescopes (MAST) at \texttt{https://mast.stsci.edu/portal/ Mashup/Clients/Mast/Portal.html}. The data in Fig.\,\ref{fig:vvori} were published by Eaton \cite{Eaton75apj}. The data in Fig.\,\ref{fig:hip} were provided by the European Space Agency (ESA) and obtained from the Centre de Donn\'ees astronomiques de Strasbourg (CDS) at \texttt{https://cdsweb.u-strasbg.fr}. The data in Fig.\,\ref{fig:corot} were obtained from the IAS \textit{CoRoT} public archive at \texttt{http://idoc-corot.ias.u-psud.fr/sitools /client-user/COROT\_N2\_PUBLIC\_DATA/project-index.html}. The data in Figs.\ \ref{fig:mr} and \ref{fig:hrd} were taken from the DEBCat database maintained by the author at \texttt{https://www.astro.keele.ac.uk/ jkt/debcat/}.}

\acknowledgments{The author apologises for anyone whose work was not included in this review: it was simply not possible to mention every relevant piece of research due to lack of space and time. I am pleased to thank Dominic Bowman, Pierre Maxted and Kre\v{s}imir Pavlovski for valuable comments on a draft of the manuscript, and the anonymous referees for their positive reports. The \textit{CoRoT} space mission, launched on 2006 December 27, was developed and is operated by the CNES, with participation of the Science Programs of ESA, ESA's RSSD, Austria, Belgium, Brazil, Germany and Spain. This paper includes data collected by the \textit{Kepler} and TESS missions and obtained from the MAST data archive at the Space Telescope Science Institute (STScI). Funding for the \textit{Kepler} mission is provided by the NASA Science Mission Directorate. Funding for the TESS mission is provided by the NASA Explorer Program. STScI is operated by the Association of Universities for Research in Astronomy, Inc., under NASA contract NAS 5–26555.}

\conflictsofinterest{The author declares no conflict of interest.}

\end{paracol}


\reftitle{References}
\externalbibliography{yes}

\end{document}